\documentclass[letterpaper,11pt]{article}
\pdfoutput=1
\usepackage{jheppub}
\usepackage{etoolbox}% http://ctan.org/pkg/etoolbox
    \makeatletter
    \patchcmd{\maketitle}{\@fpheader}{}{}{}
    \makeatother
\usepackage{slashed}
\usepackage{graphicx}
\usepackage{subfigure}
\usepackage{soul}
\usepackage{amsmath}
\usepackage{multirow}
\usepackage{hyperref}
\usepackage{epstopdf}
\usepackage{indentfirst}
\newcommand{\beq}{\begin{equation}}
\newcommand{\eeq}{\end{equation}}
\newcommand{\bea}{\begin{eqnarray}}
\newcommand{\eea}{\end{eqnarray}}

\def\m1{M_1}
\def\m2{M_2}
\def\m3{M_3}

\def\ch10{\tilde \chi^0_1}

\def\tev{\,{\rm TeV}}
\def\gev{\,{\rm GeV}}

\def\to{\rightarrow}

\newcommand{\lsim}{\mathrel{\mathop{\kern 0pt \rlap
  {\raise.2ex\hbox{$<$}}}
  \lower.9ex\hbox{\kern-.190em $\sim$}}}
\newcommand{\gsim}{\mathrel{\mathop{\kern 0pt \rlap
  {\raise.2ex\hbox{$>$}}}
  \lower.9ex\hbox{\kern-.190em $\sim$}}}

\definecolor{pink}{RGB}{255,105,180}

\def\fbi{\,{\rm fb}^{-1}}
\def\abi{\,{\rm ab}^{-1}}

\newcommand{\ee}{{$e^{-} e^{+}$}}
\newcommand{\fb}{{\,{\rm fb}}}
\newcommand{\ifb}{{\,{\rm fb}^{-1}}}

\newcommand{\mrec}{m_{\rm rec}}
\newcommand{\mee}{m_{ee}}

\newcommand{\quotes}[1]{{\scriptsize #1}}

\title{Improving Higgs coupling measurements through $ZZ$ fusion at the ILC}

\author[a]{Tao Han,}
\author[a,b]{Zhen Liu,}
\author[a]{Zhouni Qian,}
\author[a]{Josh Sayre }

\affiliation[a]{Pittsburgh Particle Physics, Astrophysics, and Cosmology Center, \\
Department of Physics and Astronomy, University of Pittsburgh, \\
3941 O'Hara Street, Pittsburgh, Pennsylvania 15260, USA}
\affiliation[b]{Theoretical Physics Department, Fermi National Accelerator Laboratory, Batavia, Illinois, 60510, USA}

\emailAdd{than@pitt.edu}
\emailAdd{zhl61@pitt.edu}
\emailAdd{jsayre@pitt.edu}
\emailAdd{zhq8@pitt.edu}

\abstract{We evaluate the $e^- e^+ \to e^- e^+ +h $ process through the $ZZ$ fusion channel at the International Linear Collider operating at $500~\gev$ and $1~\tev$ center-of-mass energies. We perform realistic simulations on the signal process and background processes. With judicious kinematic cuts, we find that the inclusive cross section can be measured to $2.9\%$ after combining the $500~\gev$ at $500~\fbi$ and $1~\tev$ at $1~\abi$ runs. A multivariate log-likelihood analysis further improves the precision of the cross section measurement to $2.3\%$. We discuss the overall improvement to model-independent Higgs width and coupling determinations and demonstrate the use of different channels in distinguishing new physics effects in Higgs physics. Our study demonstrates the importance of the $ZZ$ fusion channel to Higgs precision physics, which has often been neglected in the literature.}

\keywords{Higgs boson, ILC, Couplings, Naturalness}
\preprint{
\begin{flushright}
PITT PACC 1505\\
FERMILAB-PUB-15-101-T
\end{flushright}
}

\begin{document}
\maketitle
\flushbottom

\section{Introduction}

The discovery of a $126$ GeV Higgs boson at the LHC completes the roster of particles predicted by the Standard Model (SM). High-energy experiments now continue their search for physics beyond the Standard Model in light of this new era. A major new avenue for pursuing this search is the detailed study of the Higgs itself. While the mass of the Higgs boson is a free parameter in the SM, its couplings to other particles are dictated by the  gauge and Yukawa interactions. The observations of this particle are so far consistent with the SM expectations, but there is considerable room for new physics to reveal itself in deviations of the Higgs properties from the SM. There are also many theoretical scenarios in which such deviations would arise at a potentially detectable level. 
Hence, a precise measurement of those couplings is a key tool in establishing a departure from the SM, and in characterizing any sign of new physics which may be discovered.

The LHC will continue to accumulate a large amount of data at unprecedented energies for many years, which will improve on the current understanding of Higgs physics. It also faces certain limitations intrinsic to a hadron collider, including the uncertainty of large QCD-related backgrounds. The LHC can measure particular channels involving specific modes of production and decay in combination, and thus constrain combinations of coupling constants and the unknown width. Unfortunately, because it cannot measure a single coupling independent of the width, it cannot place strong bounds on the absolute values of couplings, nor on  the total width unless additional, model-dependent, assumptions are made \cite{Duhrssen:2004cv,Barger:2012hv,Peskin:2012we,LHCHiggsCrossSectionWorkingGroup:2012nn,Dobrescu:2012td,Han:2012rb}. Interference effects can be used to bound the width at a few times its SM value \cite{Martin:2012xc,Kauer:2012hd,Dixon:2013haa,Caola:2013yja,Campbell:2013wga,Campbell:2013una,Khachatryan:2014iha,ATLASwidth,Liebler:2015aka}. 
A ``Higgs factory" such as the International Linear Collider (ILC) has the potential to make precision measurements of Higgs physics that take advantage of the simple reconstructable kinematics and clean experimental environment. One especially appealing feature of the ILC is the ability to accurately extract the Higgs width in a model-independent manner.

The key feature of a lepton collider in making  model-independent measurements is the ability to determine the inclusive Higgs production rate.  This is done using processes such as $e^- e^+ \to h +X$ where $X$ represents additional measurable particles. Since the initial state, including longitudinal momentum, is well known we can infer the Higgs momentum without specifying the decay of the Higgs,
\beq
p_h = p_{e^-e^+} - p_X.
\eeq
This complete kinematical reconstruction allows us to discriminate the inclusive Higgs signal from background and measure the couplings of the relevant production mechanism independently of the width. Once this is done, measurements of additional specific decay channels can be used to determine the total width and the absolute values of other couplings. In a previous study we discussed this general strategy in detail \cite{Han:2013kya}. Based on available analyses the model-independent Higgs width $\Gamma_h$ can be measured 
at the level of $\delta_{\Gamma_h} \simeq 5 \%$ relative to the true width. Most of this error derives from the uncertainty on the inclusive cross section. Thus, any substantial improvement of the total width measurement depends critically on improving the precision on the inclusive cross section. Currently, the inclusive cross section sensitivity is estimated for the ``Higgsstrahlung" channel $e^-e^+ \to Zh$. The cross section for this channel is largest just above the threshold at a center-of-mass energy $\sqrt{s} \simeq 250$ GeV, where it can be measured using the $Z$ decay to electrons and muons with a relative error  $\delta \sigma^\text{inc}_{Zh} \simeq 2.6 \%$ \cite{Li:2012taa,Li:2010wu}. 
At $\sqrt{s} = 500$ GeV the Higgsstrahlung rate is substantially reduced but using hadronic decays of the $Z$ may allow one to measure the cross section at $\delta \sigma^\text{inc}_{Zh} \simeq 3 \%$~\cite{Miyamoto:2013zva}.

\begin{figure}[t]
	\centering
	\includegraphics[scale=0.4]{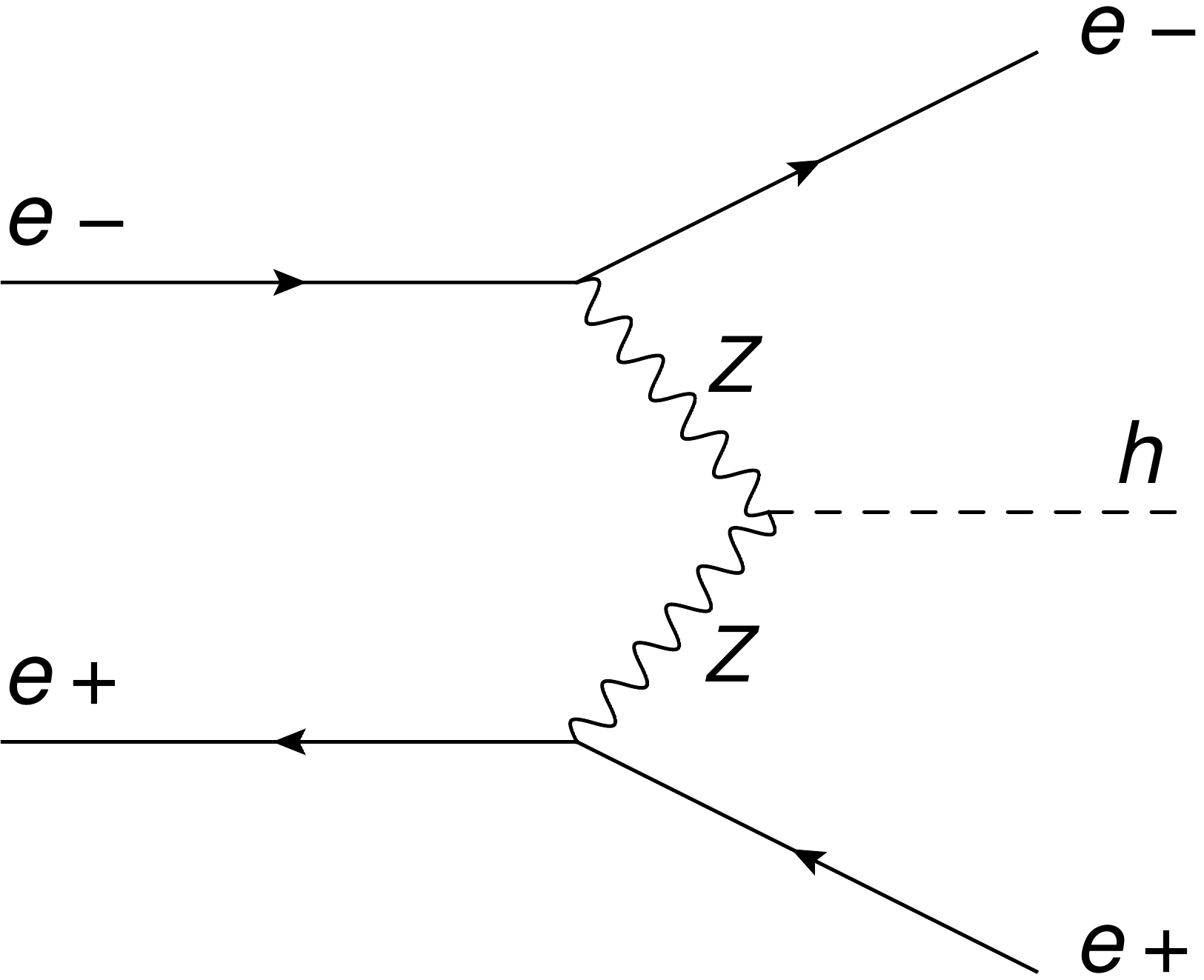}
	\caption[]{Feynman diagram of the $ZZ$ fusion signal process}
	\label{fig:zzh}
\end{figure}

\begin{figure}[t]
	\centering
	\includegraphics[scale=0.5]{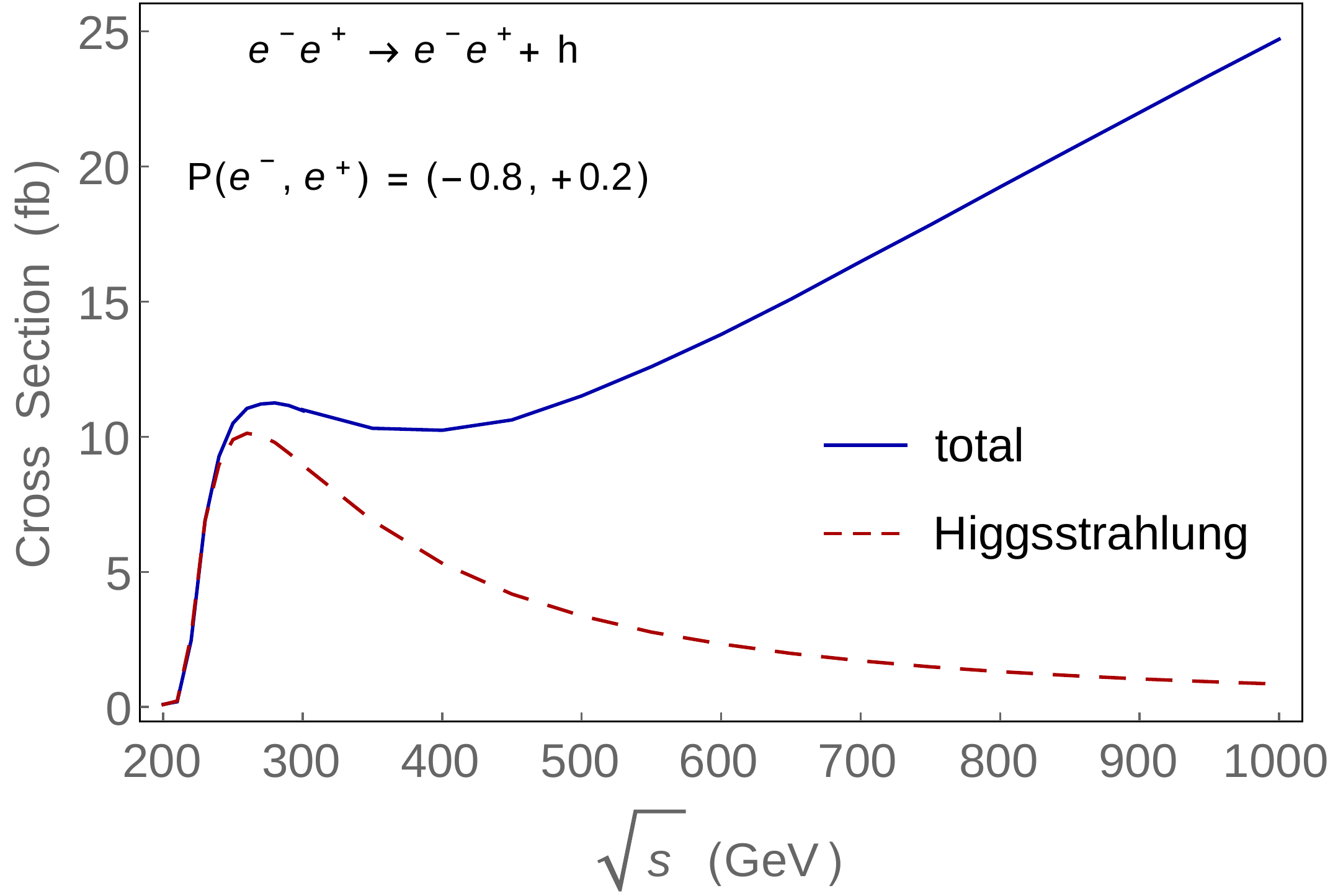}
	\caption{Total cross section (in fb) for $e^- e^+ \to e^- e^+ +h $ at ILC versus $\sqrt{s}$. The dashed curve is for Higgsstrahlung mode only.
	}
	\label{fig:xsec}
\end{figure}

Further improvements can be made by examining the alternate production mechanism of $ZZ$ fusion \cite{GunionHan,HanJiang} 
\beq 
e^-e^+ \to e^-e^+Z^* Z^* \to e^-e^+h,
\label{eq:ZZfuse}
\eeq 
as depicted in Fig.~\ref{fig:zzh}, which has often been neglected in the literature. This mode has a small rate at $250$ GeV but grows with energy as $\ln^2(s/M^2_Z)$. At $500$ GeV it already contributes roughly twice as 
much to the final state $e^-e^+h$ as the Higgsstrahlung process $Zh\to e^-e^+h$, which falls roughly as $1/s$, as can be seen in Fig.~\ref{fig:xsec}. At $1$ TeV this ratio grows to almost a factor of 20. Thus, although the Higgsstrahlung process benefits from a sharp kinematic on-shell $Z$ peak through the reconstructible final states into which the $Z$ decays, the $ZZ$ fusion channel, which features two energetic forward/backward electrons, should also be exploited to make maximal use of the high-energy reach of the ILC.

In this work we perform a fast detector simulation analysis of the inclusive $ZZ$ fusion channel measurement at $500$ GeV and $1$ TeV. We simulate the predominant backgrounds and a SM-like Higgs signal and calculate the signal sensitivity using a cut-based analysis and multivariate log-likelihood analysis. We find that with the cut-based analysis, we can reach a sensitivity on the cross section to the $2.9\%$ level. The multivariate analysis further improves the precision of the cross section measurement to $2.3\%$. 

The rest of the paper is organized as follows: In Sec.~\ref{sec:sen}, we discuss the kinematic features for identifying the signal and perform a detailed analyses for the $ZZ$ fusion process at $500$ GeV and $1$ TeV energies including backgrounds. In Sec.~\ref{sec:disc} we discuss the effects of this additional information on the model-independent Higgs width and couplings. We also illustrate the potential use of these couplings in constraining higher-dimensional operators. We summarize our results in Sec.~\ref{sec:conclusion}.
An appendix is included to address issues relating to potential signal and backgrounds with a single photon in the final state.

%%%%%%%%%%%%%%%%%%%%%%%%%%%%%%%%%%%%%%%%%%

\section{Sensitivity Analysis}
\label{sec:sen}

We consider the signal process $e^-e^+ \to e^-e^+h$ via $ZZ$ fusion as in Eq.~(\ref{eq:ZZfuse}). 
We assume that the incoming leptons are described by the nominal beam energy moving along the beam axis in the positive and negative directions respectively. Then the outgoing electrons are each characterized by a three-dimensional vector and there are six independent degrees of freedom measured in our final state. We choose the dimensionful variables to be the invariant mass of the final electron-positron pair $m_{ee}$ and the recoil mass, given by
\beq
m_{rec}^2 \equiv s -2\sqrt{s}E_{ee} +m_{ee}^2.
\eeq

The recoil mass provides the most distinct signal feature since it displays the resonance peak at the Higgs mass $m_h \simeq 126$ GeV observable on top of a continuum background. The electron-pair mass $m_{ee}$ favors a large value $m_{ee} \gtrsim 250\ (600)$ GeV at a $500\  (1000)$ GeV center-of-mass energy. This is distinct from the Higgsstrahlung mode where the pair mass is strongly peaked at the $Z$ resonance. 
Despite a broad distribution for the $ee$ pair mass in the $ZZ$ fusion, it still provides some discriminating power against the diffuse electron background.

\begin{figure}[t]
	\centering
	\includegraphics[scale=0.5]{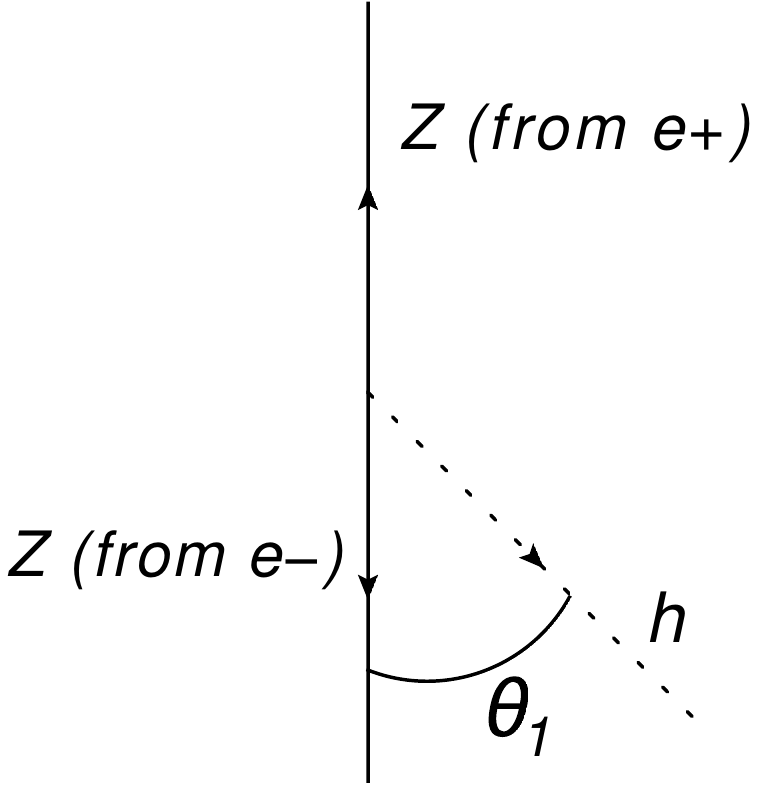} \qquad
        \includegraphics[scale=0.5]{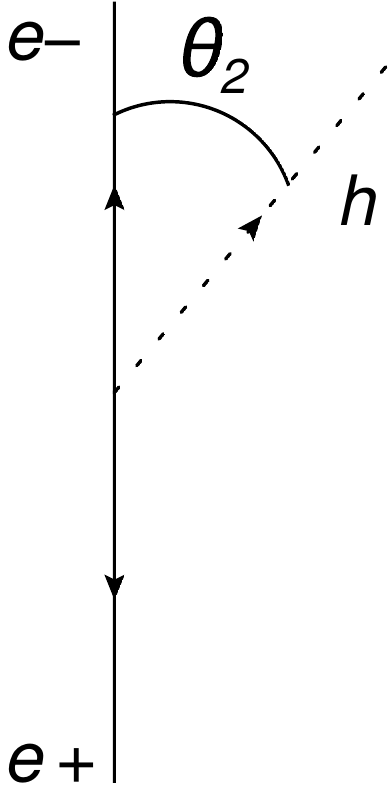} \qquad
        \includegraphics[scale=0.5]{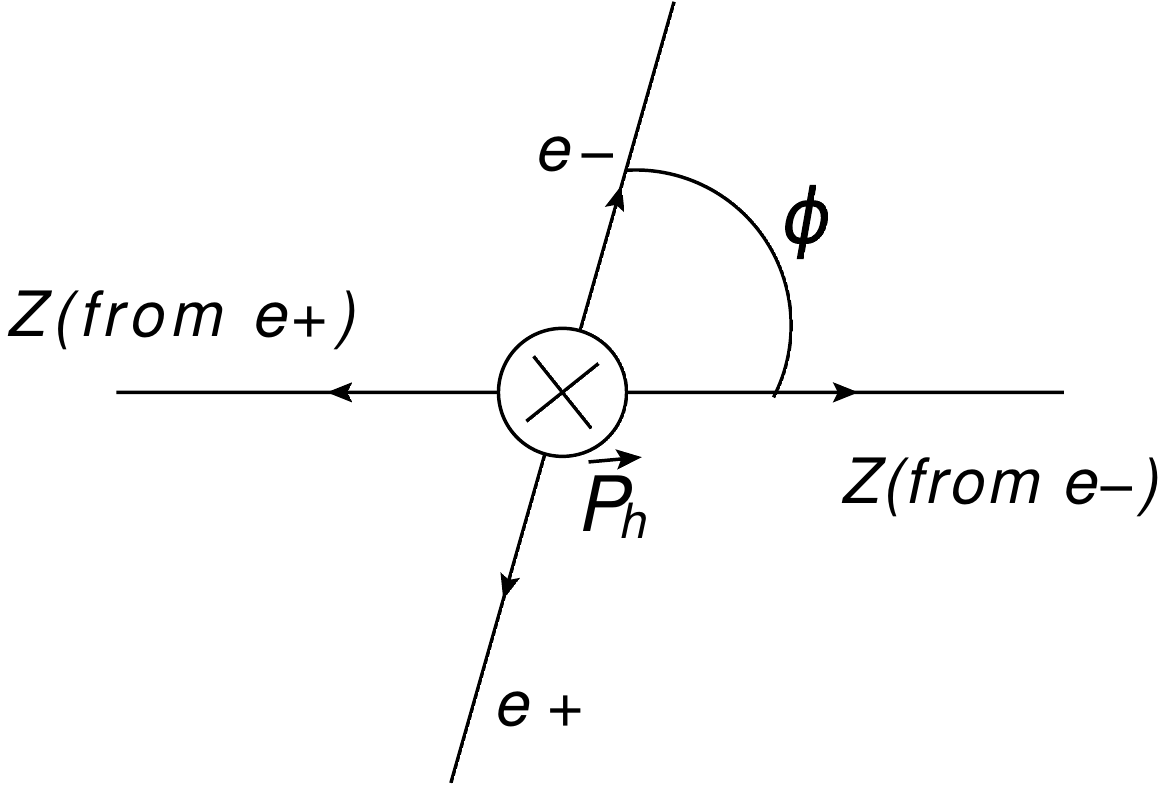}
	\caption[]{Angles $\theta_{1}$, $\theta_{2}$ and $\phi$ as defined in the text. The label 
	$e^-~(e^+)$ represents the outgoing electron (positron) and the $Z$ momentum is given by the difference between outgoing and incoming electrons (positrons). The arrows represent momentum directions.  The Higgs momentum is perpendicular to the plane in the right panel.
	}
	\label{fig:theta1}
\end{figure}

The remaining kinematic degrees of freedom can be described by four angles. One of these, the azimuthal angle of the Higgs boson around the beam axis, is irrelevant to our analysis due to the rotational symmetry of the initial state around the beam line
when the beam is not transversely polarized. The other three angles, illustrated in Fig.~\ref{fig:theta1}, are chosen as follows: $\theta_1$ is the angle between the intermediate $Z$ coming from the initial electron and the Higgs boost direction in the rest frame of the Higgs. 
$\theta_2$ is the angle between the final state electron and the Higgs boost direction in the rest frame of the outgoing $e^-e^+$ pair.
These angles take advantage of the scalar nature of the Higgs.  The distributions for $\cos\theta_1$ and $\cos\theta_2$ are rather flat since the Higgs boost direction has no preference to align with the spins of the incoming $Z$s or outgoing electrons. There is some correlation between these two angles and mild enhancement at larger $|\cos{\theta}|$, which corresponds to a more collinear configuration. This is mitigated by the relatively large virtuality of the $Z$ propagators. In contrast the most important backgrounds show much stronger correlation and peaks at high $|\cos{\theta}|$ arising from highly collinear regions of phase space which tend to dominate their production. 
The third variable, $\phi$, is defined as the angle between the plane defined by the $ZZ$ pair and the plane defined by the outgoing $e^-e^+$ pair when viewed along the Higgs boost direction. It is a measure of coplanarity. Here the signal shows a preference for small values of $\phi$, indicating coplanar emission of the outgoing $e^-e^+$ pair with the $Z$ propagators and with the incoming leptons. This strong correlation is expected since the Higgs does not carry away any spin information. The backgrounds will generally have a more complex spin structure which is not strongly coplanar. 

In practice, the outgoing electrons of our signal will tend to radiate photons, an effect we treat with showering. This radiation degrades our signal resolution. To ameliorate this, nearby photons are clustered according to a recombination algorithm and identified with a single electron as described in detail in the next section.

Given our inclusive signal process, the backgrounds are of the form $e^-e^+ \to e^-e^+ X$.  
Obviously, the single photon radiation $X=\gamma$ arising from the Bhabha scattering is by far the largest. Although the majority of  events should be removed by the requirement of a large recoil mass $m_X$, 
beamstrahlung and the effects of the initial-state radiation (ISR), as well as the final-state radiation (FSR), will produce additional largely collinear photons. This generates  a long tail in the recoil mass spectrum due to unobserved photons, mainly along the beam pipe. 
To keep this class of backgrounds under control, we introduce a cut on the transverse momentum $p_T$ of the outgoing $e^-e^+$ pair. Photons which are lost down the beam pipe should only contribute small $p_T$ differences to the observed final state. Thus the final state $e^-e^+$ intrinsically has no $p_T$ as long as collinear photons from final-state showering are correctly regrouped with the electrons. The signal, in contrast, has a nonzero $p_{T}$ from the recoiling Higgs.

This leaves a background from $e^-e^+ \gamma$ where the extra photon is not close enough to either electron to be grouped with it by the clustering algorithm. 
We find it most convenient to simply veto events, in addition to the $e^-e^+$ pair, with a single isolated photon
\beq
\label{eq:gammacut}
E_\gamma > 10~\gev,\quad \theta_\gamma > 6^\circ,
\eeq
where $\theta_\gamma$ is the polar angle with respect to the beam.
The effectiveness of this cut is illustrated in Table~\ref{tab:pcut} for the $500\gev$ and $1\tev$ runs. (See the next section in Tables~\ref{tab:500cuts} and \ref{tab:1000cuts} for numerical definitions of the cuts.) Simple cuts on invariant mass and $p_T$ reduce the $e^-e^+ \gamma$-induced background by 3 orders of magnitude but it remains $30$ times larger than our signal. However the single photon veto reduces this by more than $90\%$. 
\begin{table}[tb]
\centering
\begin{tabular}{|c|c|c|c|c|}
\hline
 Cuts ($\fb$) & Generator level & $ m_\text{rec},~m_{ee} $ & $p^{}_{T(ee)}$ & Veto isolated single $\gamma$ \\ \hline
\ee $h$ ($500\gev$) & $11.5$ & $4.11$ & $3.48$ & $3.48$\\ \hline
\ee $ \gamma$ ($500\gev$) & $165000$ & $317$ & $67.2$ & $1.32$\\ \hline
\hline
\ee $h$ ($1\tev$)& $24.1$ & $9.75$ & $8.49$ & $8.18$\\ \hline
\ee $ \gamma$ ($1\tev$)& $175000$ & $1570$ & $344$ & $4.73$\\ \hline
\end{tabular}
\caption[]{Cross section ($\fb$) for signal  \ee +$h$ and background \ee$\gamma$ after sequence of cuts. The cuts are specified in Table~\ref{tab:500cuts} and Table~\ref{tab:1000cuts} for the $500\gev$ and $1\tev$ case respectively. 
}
\label{tab:pcut}
\end{table}

In principle this affects our inclusiveness. However, the Standard Model processes which could produce such a signal, such as $h\to \gamma \gamma$ (where one photon is lost down the beam pipe) and $h \to Z \gamma$, constitute branching fractions of $2.3 \times 10^{-3}$ and $1.6 \times 10^{-3}$ respectively. As will be seen, the ultimate precision for the inclusive Higgs production measurement is at the $\sim 2 \%$ level so that Higgs decays to $\gamma \gamma$ or $Z \gamma$ would have to be enhanced by more than an order of magnitude compared to the Standard Model to be seen in the model-independent inclusive measurement. Any such large signal enhancements will be seen at the LHC, to the extent that they are not already excluded by current results. See the Appendix for further discussion.

After these cuts some background can remain due to poorly measured final-state particles. Particularly at $1$ TeV center-of-mass energies, errors on the detected momentum of the final state can sometimes fake a recoil mass and a high $p_T$ that passes our other cuts. This is necessarily an issue to be determined in detail by experimentalists when working with an actual machine and is only parameterized by assumptions on detector smearing and efficiency in our simulation. We find that badly measured states are typically associated with very high-energy photons. Either these photons are not detected at all due to imperfect calorimeter efficiency, or they are reported but with significant error on their transverse momenta. Mismeasured low-energy photons will not usually cause a big enough error to satisfy our previous cuts. Thus it is useful to veto events with very high-energy detected photons, which are relatively rare in the signal.

Again, one may worry about introducing a bias against photons from Higgs decay, but this problem can be addressed. When an event has a high-energy photon we first boost it into the rest frame of the Higgs, as determined by the momentum of the outgoing lepton pair. If the photon's energy in the Higgs frame is less than half the Higgs mass, then it potentially comes from a Higgs decay, and we do not subject it to the high-energy veto. Thus only events with  ``eligible'' photons, $\gamma^*$, which could not have come from the Higgs decay, are cut. 

\subsection{Simulation framework}
To estimate the expected number of events and derive the sensitivity reach at a given energy and luminosity we use the ILC \quotes{WHIZARD} setup provided through the detector simulation package \quotes{SGV}3 \cite{Berggren:2012ar}. Beam profiles for several energies have been generated by GuineaPIG~\cite{Schulte:1999tx}, which includes effects from beamstrahlung and ISR. These profiles are interfaced with \quotes{WHIZARD} 1.95~\cite{Kilian:2007gr} to generate parton-level samples. The parton-level samples are then passed to \quotes{PYTHIA} which performs showering and hadronization to final-state particles~\cite{Sjostrand:2006za}. \quotes{SGV} is a fast detector simulation which has been found to agree well with full simulation results.

To avoid collinear and soft divergences, at the parton level we require that 
the energy of a final state photon be greater than $10~\gev$,
and that the invariant masses of final lepton-antilepton pairs and of lepton-photon pairs be greater than $4$ GeV. We also require that the invariant mass of a final-state (anti)electron with an initial (anti)electron, or of a final photon with an initial lepton, be greater than $4$ GeV. More collinear photons will be generated via the showering routines in \quotes{PYTHIA}.

After simulating tracking and calorimeter hits, \quotes{SGV} attempts to identify charged and neutral particles and groups these into jetlike objects according to a sequential recombination algorithm. We use the JADE algorithm, which defines a distance between objects
\begin{align}
y_{ij} \equiv \frac{2 E_i E_j(1 - \cos{\theta_{ij}})}{E_\text{vis}^2},
\end{align}
where $E_i$ and $E_j$ are the energies of two objects and $E_\text{vis}$ is the total seen energy of the event. Nearby objects are merged into subjets until all subjets are separated by $y_{ij} > 0.01$. 

In selecting our observables we first identify the two highest-energy electron/positron tracks in an event and discard it if there are fewer than two detected (anti)electrons. We also require that these particles have opposite signs. If nearby calorimeter hits included in the subjet which contains the track are only identified as photons, then we use the jet momentum and energy for our 
reconstructed lepton. If the subjet contains any particles identified as hadrons then we use only the track momentum in order to minimize cases where hadron jets overlap with the recoiling electrons.
For the purposes of the isolated photon cut described above, we define an isolated photon as a jet object which contains only photons and no charged tracks or hadronic calorimeter hits.

In the case of pure photon plus electron/positron backgrounds we simulate both $e^-e^+ \to e^-e^+ \gamma$ and $e^-e^+ \to e^-e^+ \gamma \gamma$ at the matrix element level. After showering there is some overlap in the signals described by these two processes. In the spirit of matching calculations done for hadron colliders we discard events from $e^-e^+ \to e^-e^+ \gamma$ which produce two isolated photons  after the clustering procedure. 

\subsection{$500$ GeV analysis}

We proceed with a sensitivity analysis for the ILC running at a $500~\gev$ center-of-mass energy. We apply an initial beam polarization of $-0.8$ for the electron and $+0.3$ for the positron, following the ILC technical design report \cite{Baer:2013cma}. We first perform a purely cut-based analysis with the cuts listed in Table~\ref{tab:500cuts}. $E_\gamma^*$ represents only photon hits with energy greater than $65$~GeV in the rest frame of the Higgs.

\begin{table}[t]
\centering
\begin{tabular}{|c|c|}
\hline
 & $122~\gev < \mrec < 145~\gev$ \\
 & $ 110~\gev < \mee < 370~\gev$ \\
Cut 1 & $p_{T(ee)} > 40~\gev$ \\
& veto 1 isolated photon \\
&$E_{\gamma}^* < 200$~\gev \\
\hline
Cut 2
& $\phi < 1.5$ \\ \hline
\end{tabular}
\caption{Cuts applied at ILC $500$ GeV.}
\label{tab:500cuts}
\end{table}

\begin{figure}[]
\centering 
\includegraphics[width=.45\textwidth]{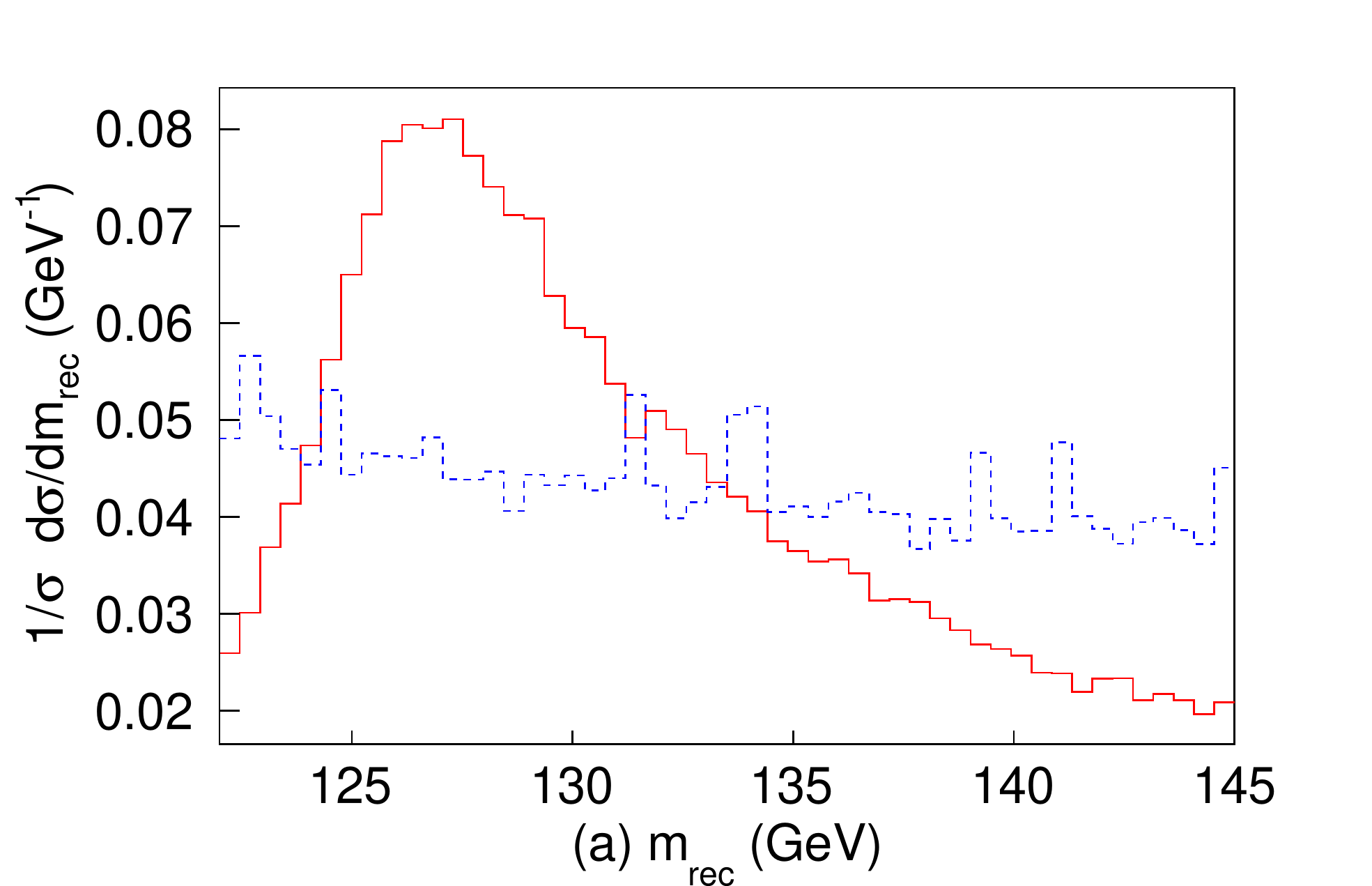}
~~
\includegraphics[width=.45\textwidth]{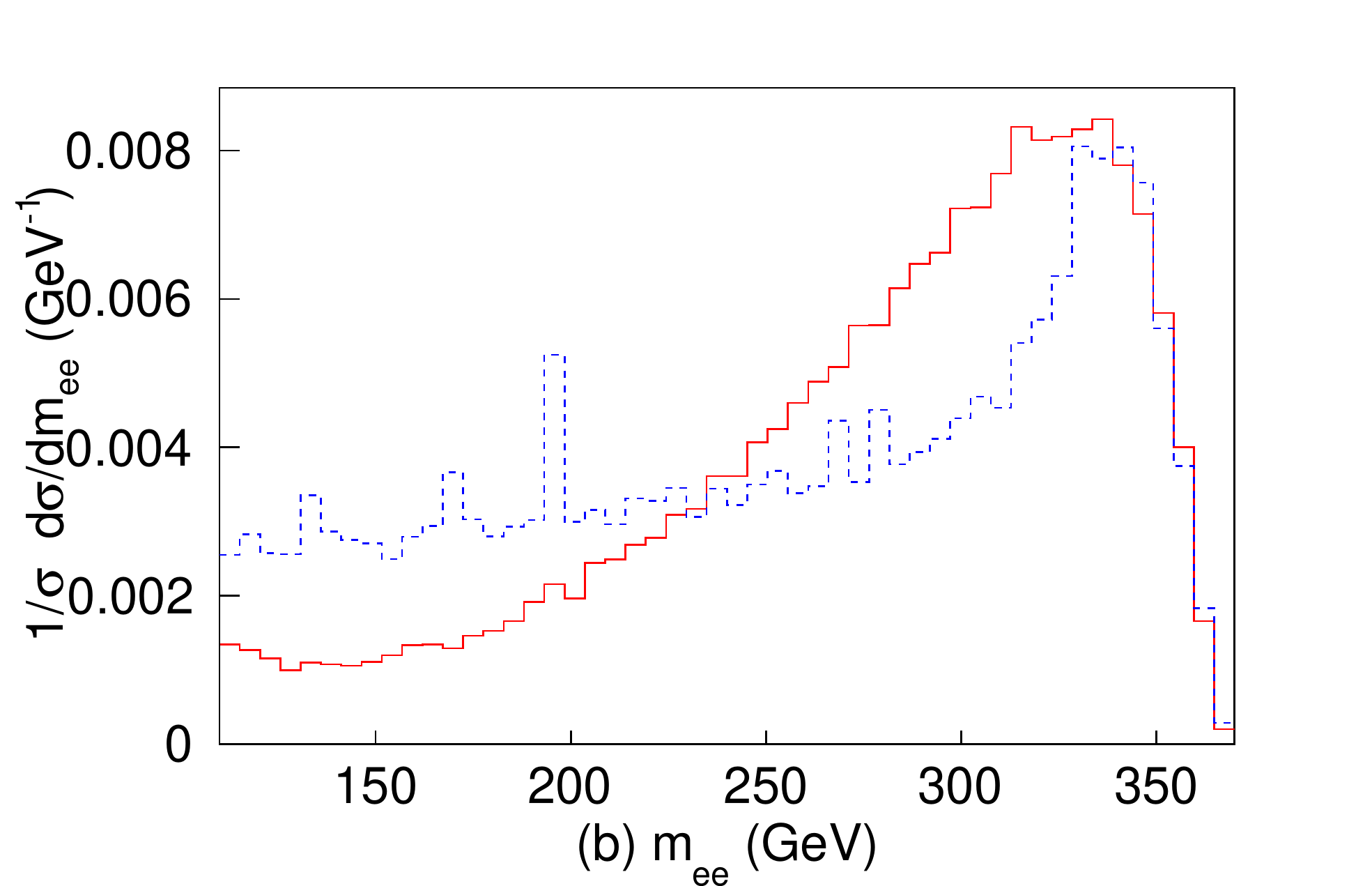}
\\
\includegraphics[width=.45\textwidth]{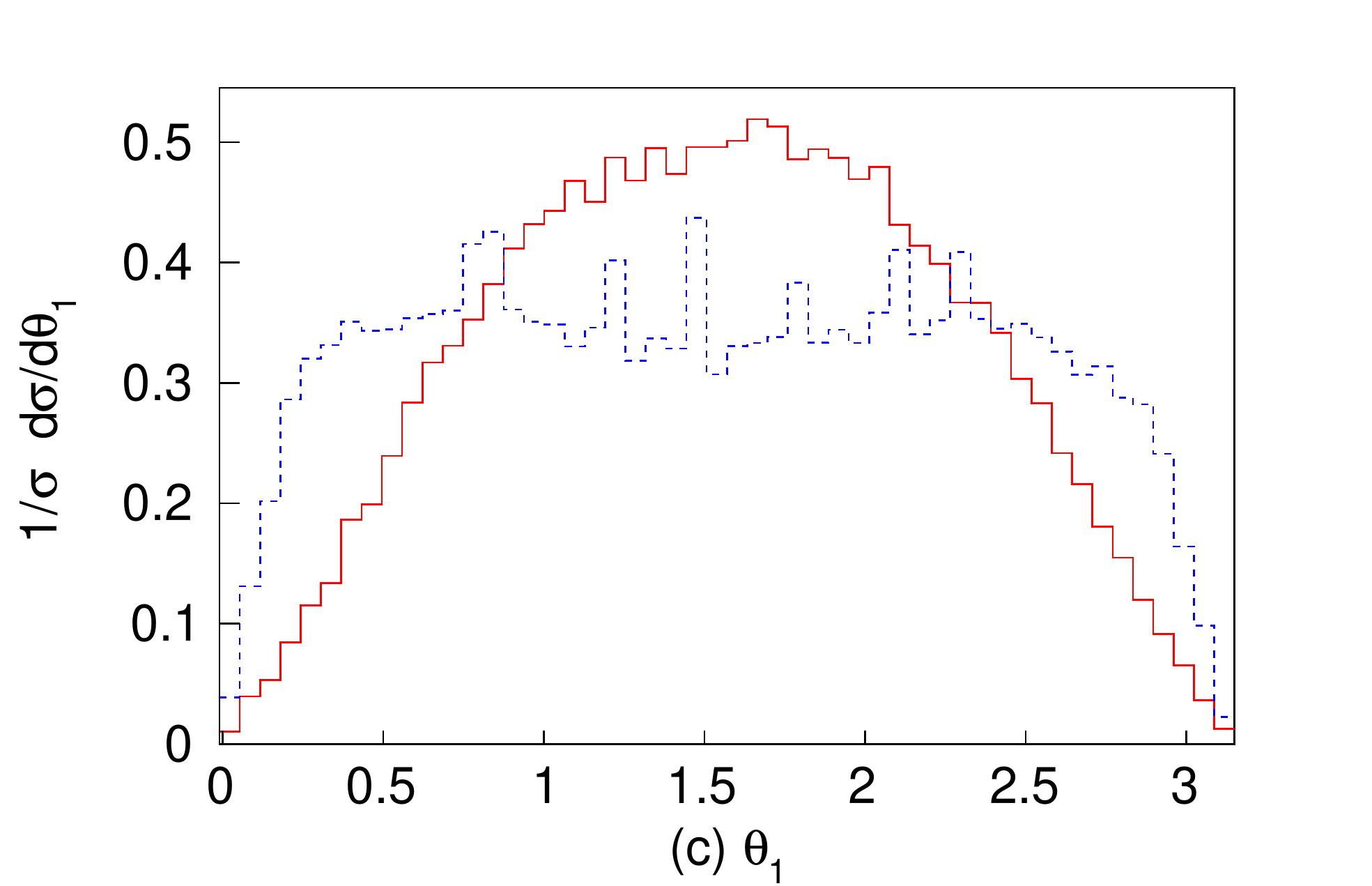}
~~
\includegraphics[width=.45\textwidth]{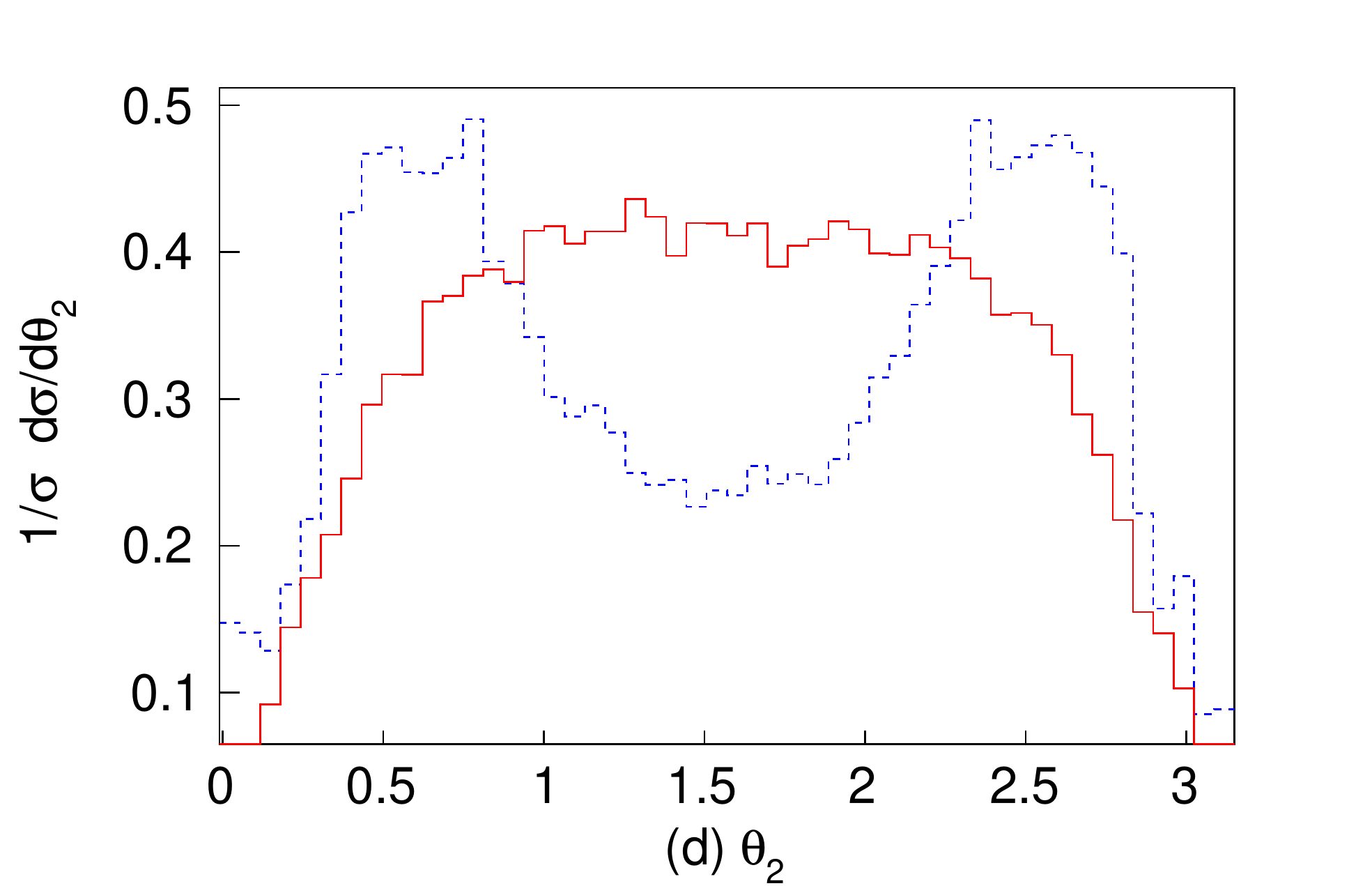}
\\
\includegraphics[width=.45\textwidth]{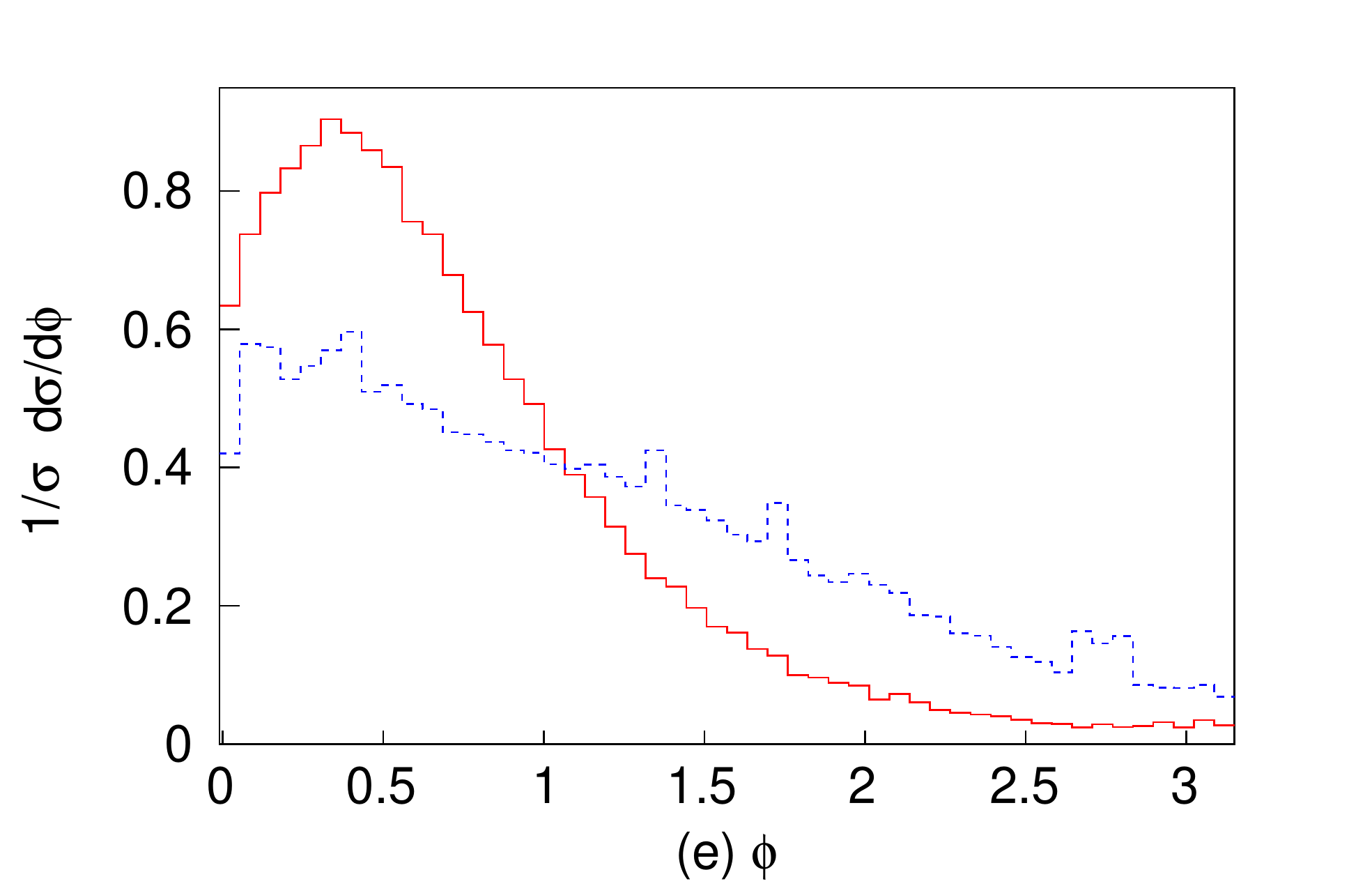}
\caption[]{Comparison of signal (solid red) and total background (dashed blue) distributions for variables (a) $m_{rec}$, (b) $m_{ee}$, (c) $\theta_1$, (d) $\theta_2$ and (e) $\phi$ at $\sqrt{s}=500~\gev$.
Cut 1 in Table~\ref{tab:500cuts} is applied.  For clarity, both signal and background distributions are normalized to unity.
 }
 \label{fig:500dist}
\end{figure}

Figure~\ref{fig:500dist} displays the signal and background distributions in $m_\text{rec}$, $m_{ee}$ and the three angular variables, after applying Cut 1 as listed. As can be seen, the 
angular variables show considerable distinction from the background which can be used to enhance our sensitivity. Cut 2 acts on these angles.

For this analysis we define the signal sensitivity according to the statistical $1\sigma$ relative error on the signal,
\begin{equation}
 \frac{\delta \sigma}{\sigma} = \frac{\sqrt{N_s+N_b}}{N_s},
\end{equation}

where $N_{s,b} = L\sigma_{s,b}$ are the expected number of signal and background events after cuts respectively. We assume the integrated luminosity $L=500\ \text{fb}^{-1}$ at this energy.
The statistical significance is then inversely related to the signal sensitivity as 
$N_s/\sqrt{N_s+N_b}$. The effect of our cuts on the cross section for signal and background processes is given in Table~\ref{table:cuts_500}. 

%where $N_s$ and $N_b$ are the expected number of signal and background events after cuts respectively. The statistical significance is then inversely related to the signal sensitivity as 
%$N_s/\sqrt{N_s+N_b}$. We assume $500\ \text{fb}^{-1}$ 
%of integrated luminosity at this energy. The effect of our cuts on the cross section for signal and background processes is given in Table~\ref{table:cuts_500}. 

%%% table in cross section (fb)
\begin{table}[t]
\centering
\begin{tabular}{|c|c|c|c|}
\hline
Process & Generator level (fb) & Cut 1 (fb) & Cut 2 (fb) \\ \hline
$ee \to eeh$(Signal) &$11.5$&$3.48$ & $3.11$ \\ \hline
$ee \to ee\nu_e\nu_e$ & $659$&$23.9$ & $16.0$ \\ \hline
$ee \to ee\nu_{\mu,\tau}\nu_{\mu,\tau}$ &$78.6$& $1.02$ & $0.70$ \\ \hline
$ee \to eeqq$ &$1850$& $9.33$ & $6.88$ \\ \hline
$ee \to eell$ &$4420$& $5.18$ & $4.42$ \\ \hline
$ee \to ee\gamma\gamma$ &$1640$& $1.18$ & $0.60$ \\ \hline
$ee \to ee\gamma$ &$165\,000$& $1.32$ & $0.66$ \\ \hline
Total background &$174\,000$& $41.9$ & $29.2$ \\ \hline
$\delta\sigma/\sigma$ & $\cdots$ & $8.7\%$ & $8.2\%$ \\ \hline
\end{tabular}
\caption[]{\label{table:cuts_500}
Cross sections for signal and background processes at
ILC $500~\gev$.}
\end{table}

%%% table in # events %%% commented out
%\begin{table}[t]
%\centering
%\begin{tabular}{|c|c|c|c|}
%\hline
%Process & Generator Level & Cut1 & Cut2 \\ \hline
%$ee \to eeh$(Signal) &$5735$&$1737$ & $1553$ \\ \hline
%$ee \to ee\nu_e\nu_e$ & $329475$&$11946$ & $7985$ \\ \hline
%$ee \to ee\nu_{\mu,\tau}\nu_{\mu,\tau}$ &$39283$& $508$ & $348$ \\ \hline
%$ee \to eeqq$ &$925100$& $4466$ & $3251$ \\ \hline
%$ee \to eell$ &$2211750$& $1935$ & $1649$ \\ \hline
%$ee \to ee\gamma\gamma$ &$820250$& $24$ & $21$ \\ \hline
%$ee \to ee\gamma$ &$82532700$& $578$ & $330$ \\ \hline
%Total Background &$86858578$& $173717$ & $15417$ \\ \hline
%$\delta\sigma$ &13.4\%& $8.4\%$ & $7.9\%$ \\ \hline
%\end{tabular}
%\caption[]{Expected number of events for signal and background at a $500\gev$
%ILC with $500\ifb$ of data.}
%\label{table:cuts_500}
%\end{table}

We find that this cut-based analysis can measure the inclusive $ZZ$ fusion signal to a relative error of $~ 8\%$.  
At this energy the dominant background after our cuts is $e^- e^+ \nu_e \bar{\nu_e}$, over 80\% of which is from the process $e^- e^+ \to W^- W^+$. The large cross section of $e^- e^+ \to W^- W^+$ is favored by the beam polarization we have used at $500$ GeV ILC. It is possible to reduce this background with a polarization that favors right-handed electrons; however, this also reduces the signal and we do not find any significant gain in sensitivity with the reversed polarization. 
It is possible to enhance sensitivity with an analysis that is sensitive to shape and to correlations between variables. This is particularly useful when the signal and background display distinct features which are not sharp enough to be efficiently cut on, as in Fig.~\ref{fig:500dist}. 

%%%%%%%%%%%%%%%%%%%%%%%%%%%%%%%%%%%%%%

\subsection{1-TeV analysis}
\label{sec:1TeV}

\begin{table}[t]
\centering
\begin{tabular}{|c|c|}
\hline
 & $95~\gev < \mrec < 300~\gev$ \\
 & $ 500~\gev < \mee < 870~\gev$ \\
Cut 1 & $p_{T(ee)} > 50~\gev$ \\
& veto 1 isolated photon \\
&$E_{\gamma}^* < 200$~\gev \\ \hline
Cut 2& $0.14 < \theta_2 < 3.0$ \\
& $\phi < 1.5 $ \\ \hline
\end{tabular}
\caption[]{\label{tab:1000cuts}
Cuts applied at ILC $1$ TeV.}
\end{table}

\begin{figure}[]
\centering 
\includegraphics[width=.45\textwidth]{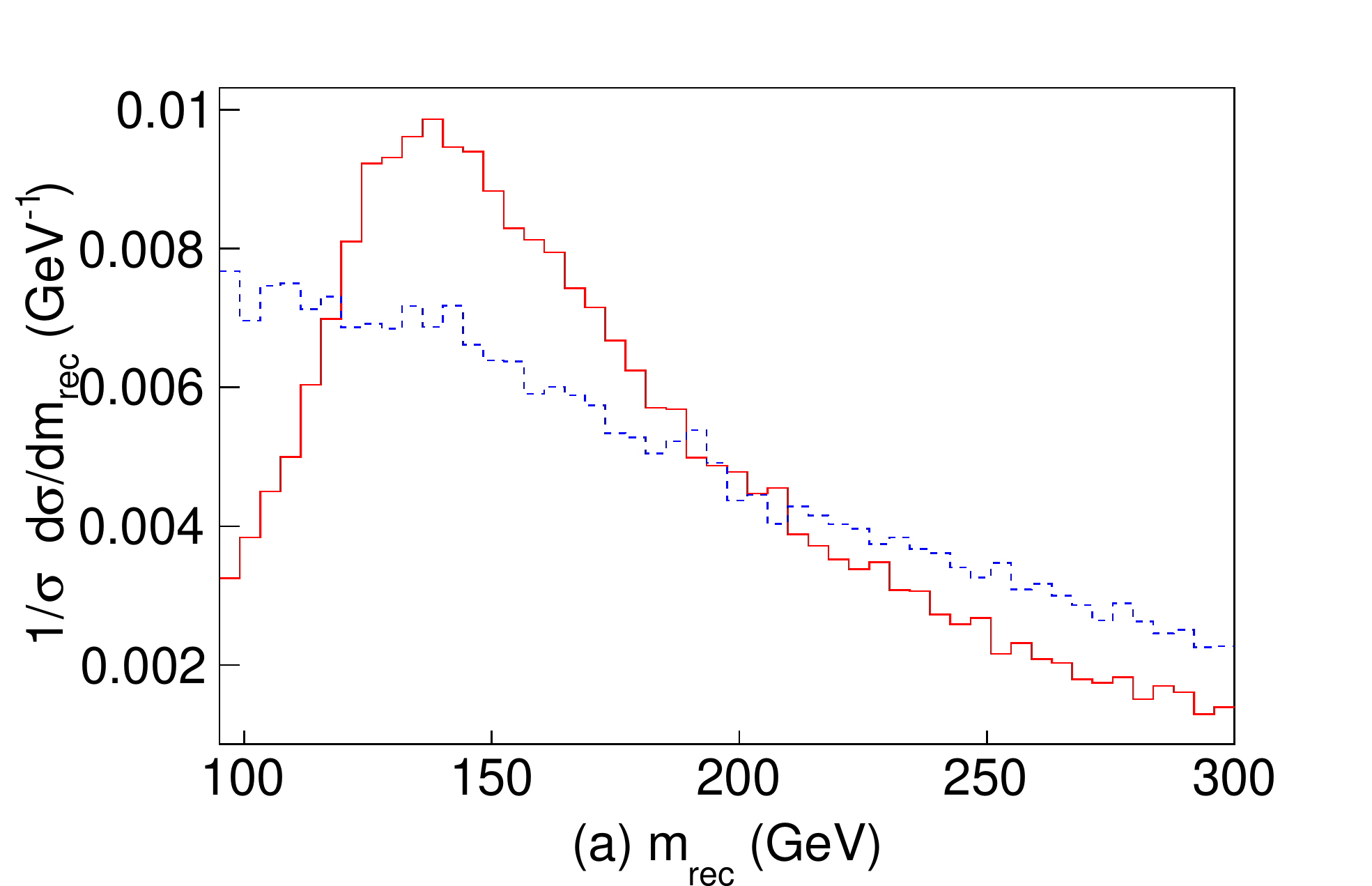}
~~
\includegraphics[width=.45\textwidth]{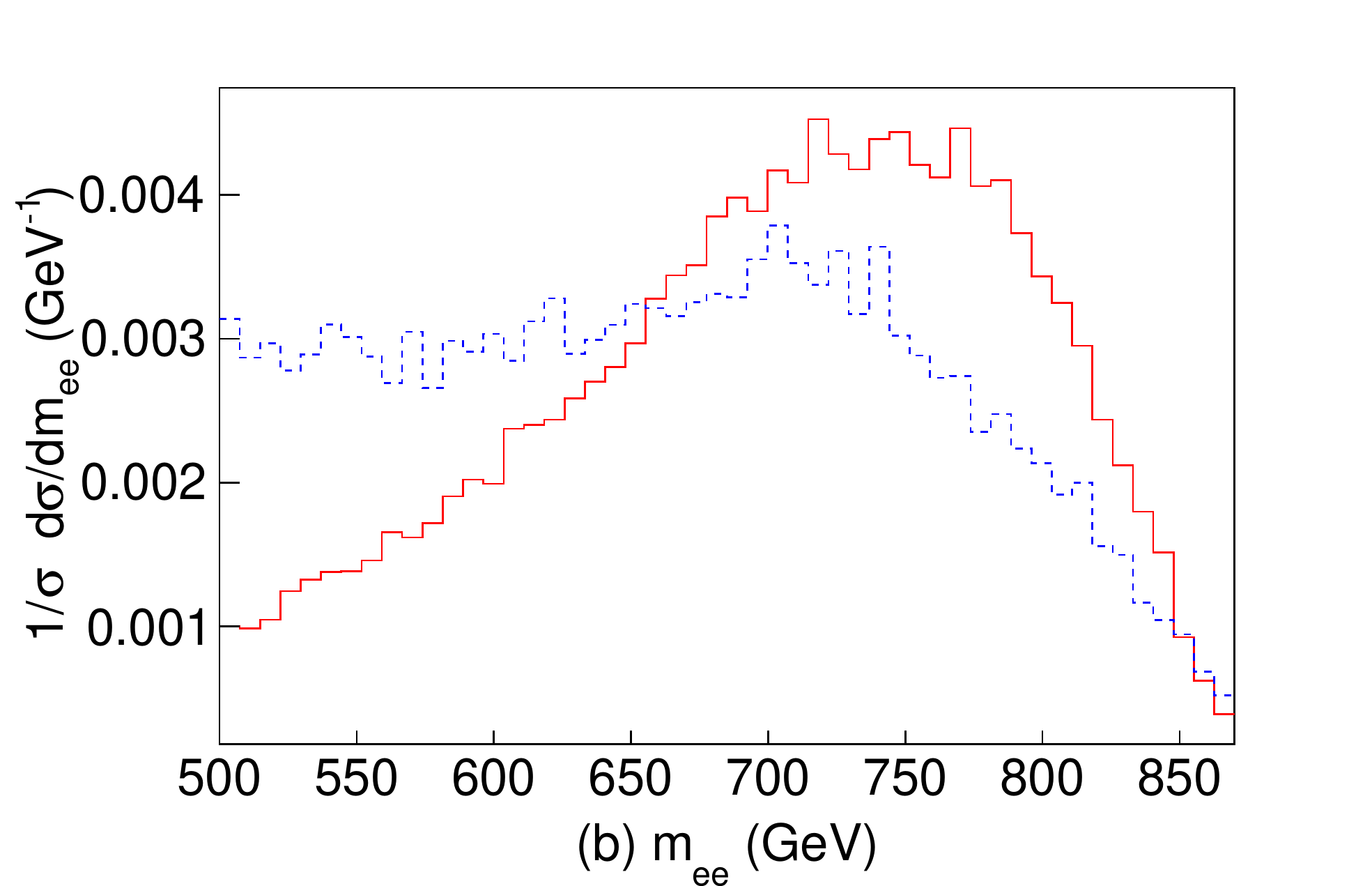}
\\
\includegraphics[width=.45\textwidth]{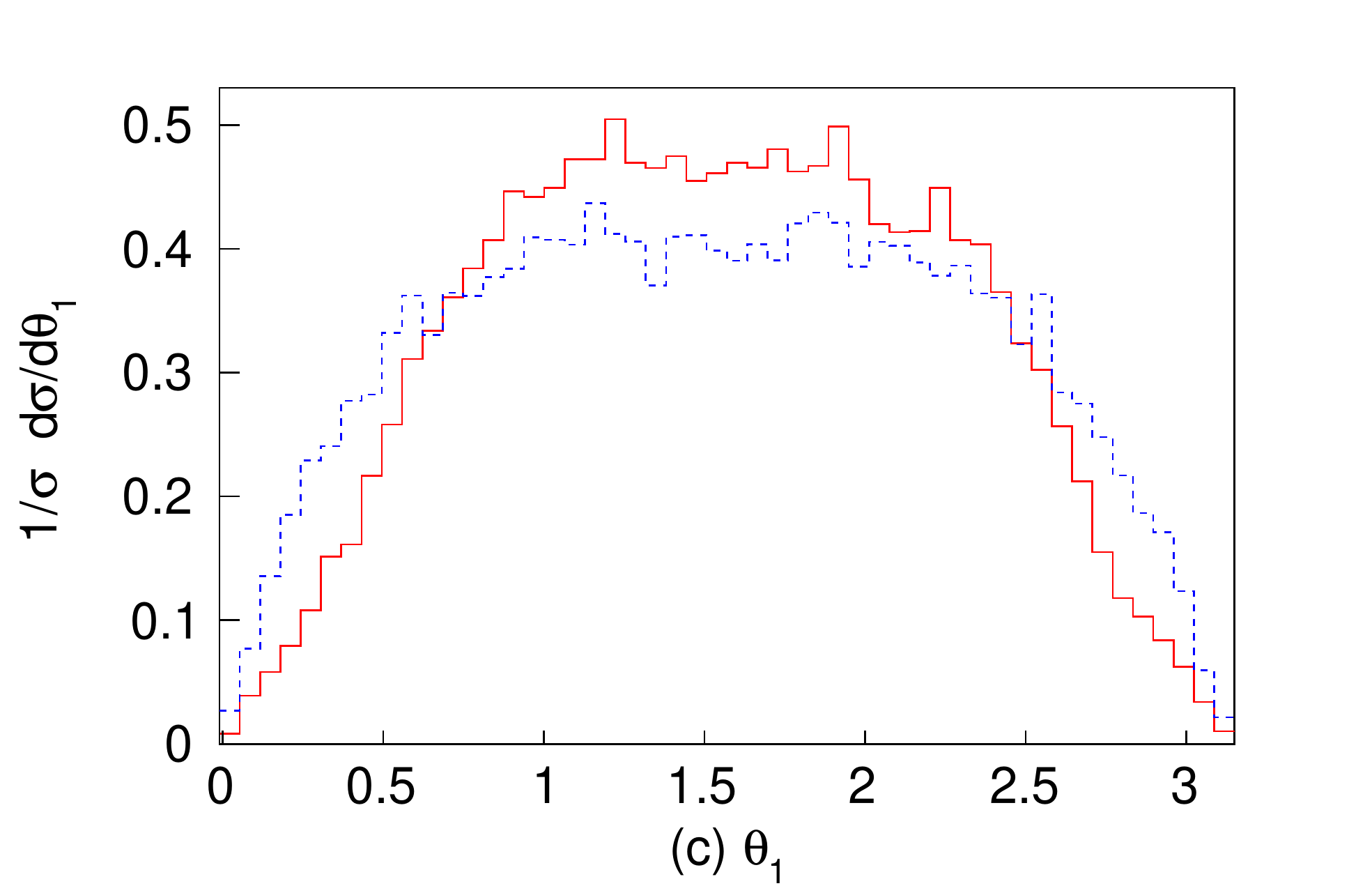}
~~
\includegraphics[width=.45\textwidth]{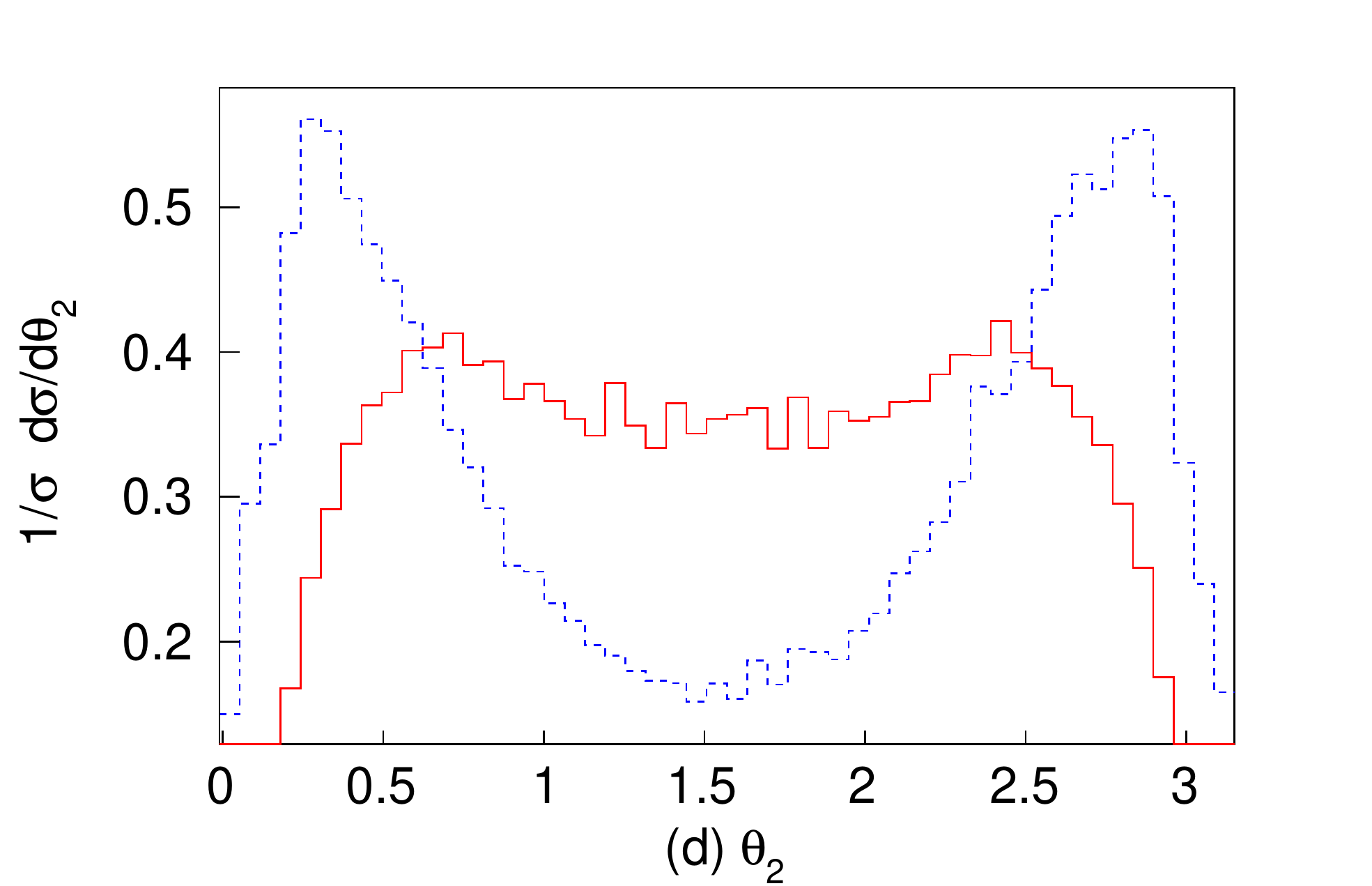}
\\
\includegraphics[width=.45\textwidth]{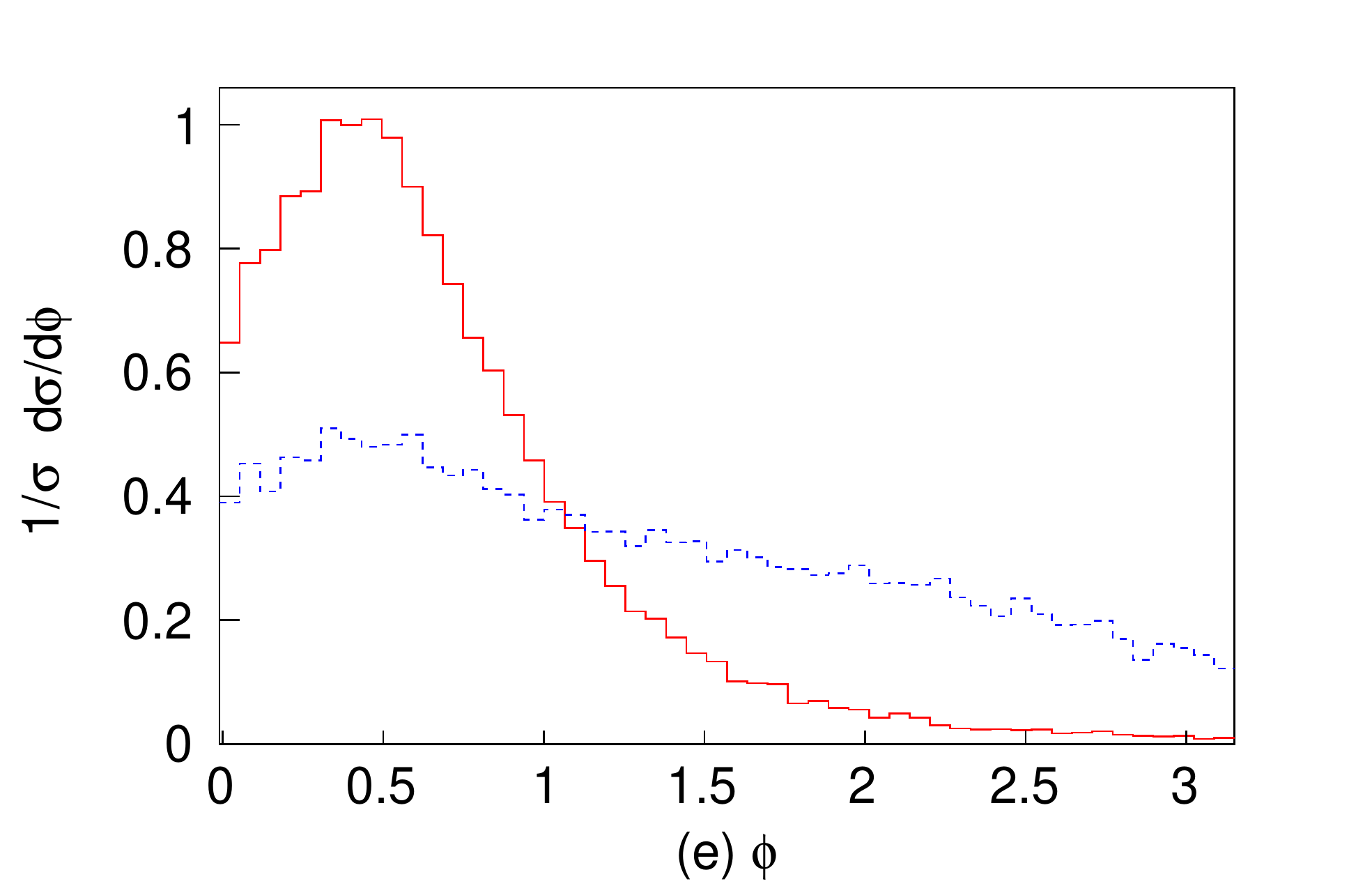}
\caption[]{Comparison of signal (solid red) and total background (dashed blue) distributions for variables $m_\text{rec}$, $m_{ee}$, $\theta_1$, $\theta_2$ and $\phi$ at $\sqrt{s}=1$ TeV.
Cut 1 in Table~\ref{tab:1000cuts} is applied.  For clarity, both signal and background distributions are normalized to unity.
}
\label{fig:1000dist}
\end{figure}

\begin{table}[tb]
\centering
\begin{tabular}{|c|c|c|c|}
\hline
Process & Generator level (fb) & Cut 1(fb) & Cut 2(fb) \\ \hline
$ee \to eeh$(Signal) &$24.1$&$8.18$ & $7.52$ \\ \hline
$ee \to ee\nu_e\nu_e$ & $978$&$31.5$ & $17.2$ \\ \hline
$ee \to ee\nu_{\mu,\tau}\nu_{\mu,\tau}$ &$93.9$& $3.24$ & $1.64$ \\ \hline
$ee \to eeqq$ &$2830$& $24.1$ & $13.6$ \\ \hline
$ee \to eell$ &$6690$& $13.7$ & $10.8$ \\ \hline
$ee \to ee\gamma\gamma$ &$3180$& $2.68$ & $1.10$ \\ \hline
$ee \to ee\gamma$ &$175\,000$& $4.73$ & $2.28$ \\ \hline
Total background &$189\,000$& $80.0$ & $46.6$ \\ \hline
$\delta\sigma/\sigma$ & $\cdots$ & $3.6\%$ & $3.1\%$ \\ \hline
\end{tabular}
\caption[]{\label{table:cuts_1000}
Cross sections for signal and background processes at
ILC $1$ TeV with $1000$ $\ifb$ of integrated luminosity.}
\end{table}

We next extend our analysis to a 1 TeV center-of-mass energy with $1000~\ifb$ integrated luminosity. The polarization is assumed to be $(-0.8,\ +0.2)$ as suggested by the Snowmass Higgs report \cite{Dawson:2013bba}. The $ZZ$ fusion process is enhanced with increased center-of-mass energy. However, due to radiation from the energetic $e^-$ and $e^+$, the Higgs mass peak in the $\mrec$ distribution is much more smeared than in the $500$ GeV case, and photon radiation backgrounds become more significant. The angular variables $\theta_2$ and $\phi$ show greater distinctions between signal and background.
To maximize significance we apply cuts as listed in Table~\ref{tab:1000cuts}.

Figure~\ref{fig:1000dist} compares the signal and total background distributions after Cut 1.
Table~\ref{table:cuts_1000} shows the expected cross sections after Cut 1 and Cut 2. Despite the degradation of the recoil mass peak we gain significance from enhanced statistics and a somewhat improved signal-to-background ratio. The cut-based analysis can reach a sensitivity of $3.1\%$. 

\subsection{Multivariate log-likelihood analyses}
\label{sec:LL}

\begin{figure}[t]
	\centering
	\includegraphics[scale=0.36]{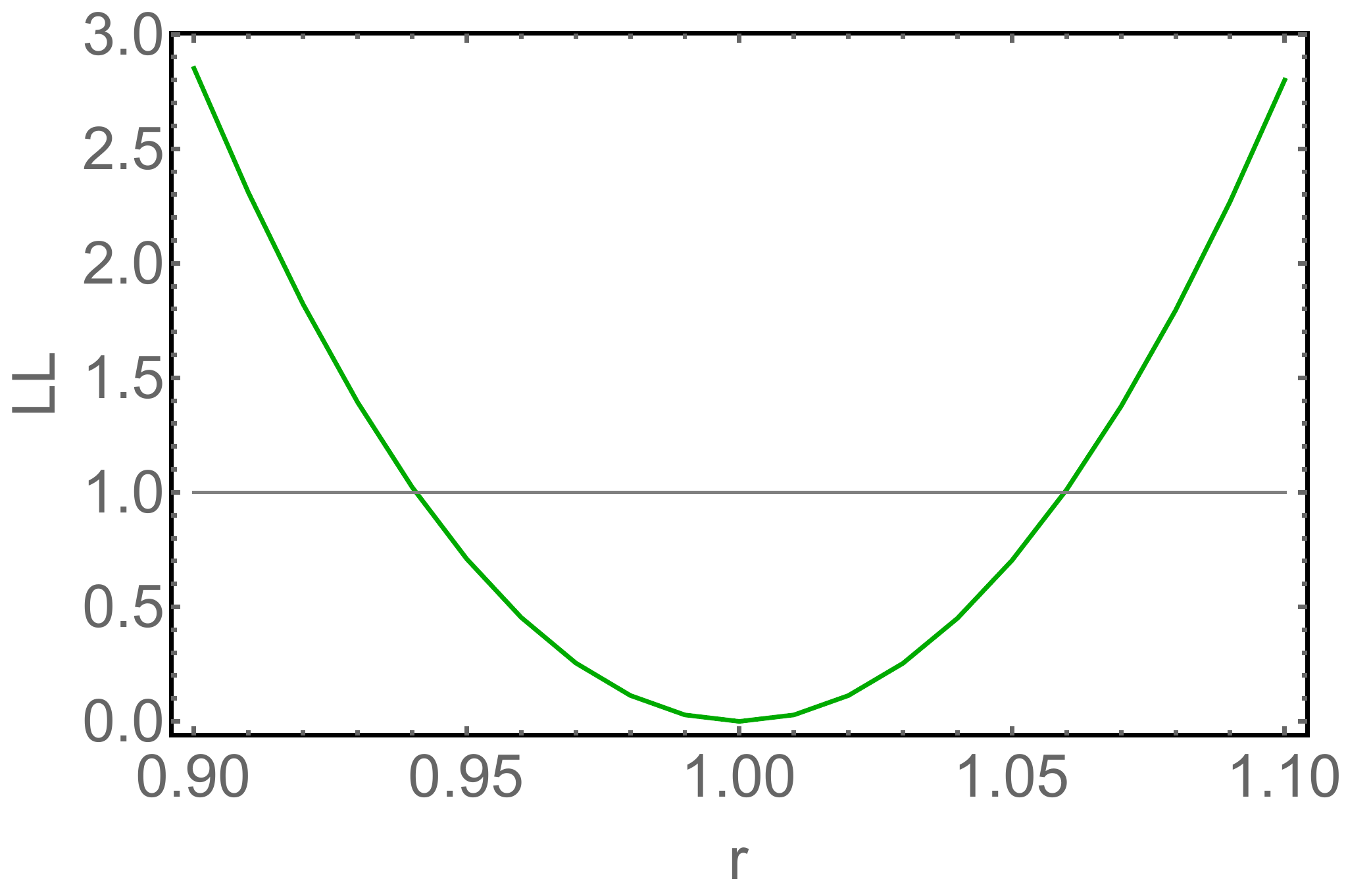}
		\includegraphics[scale=0.36]{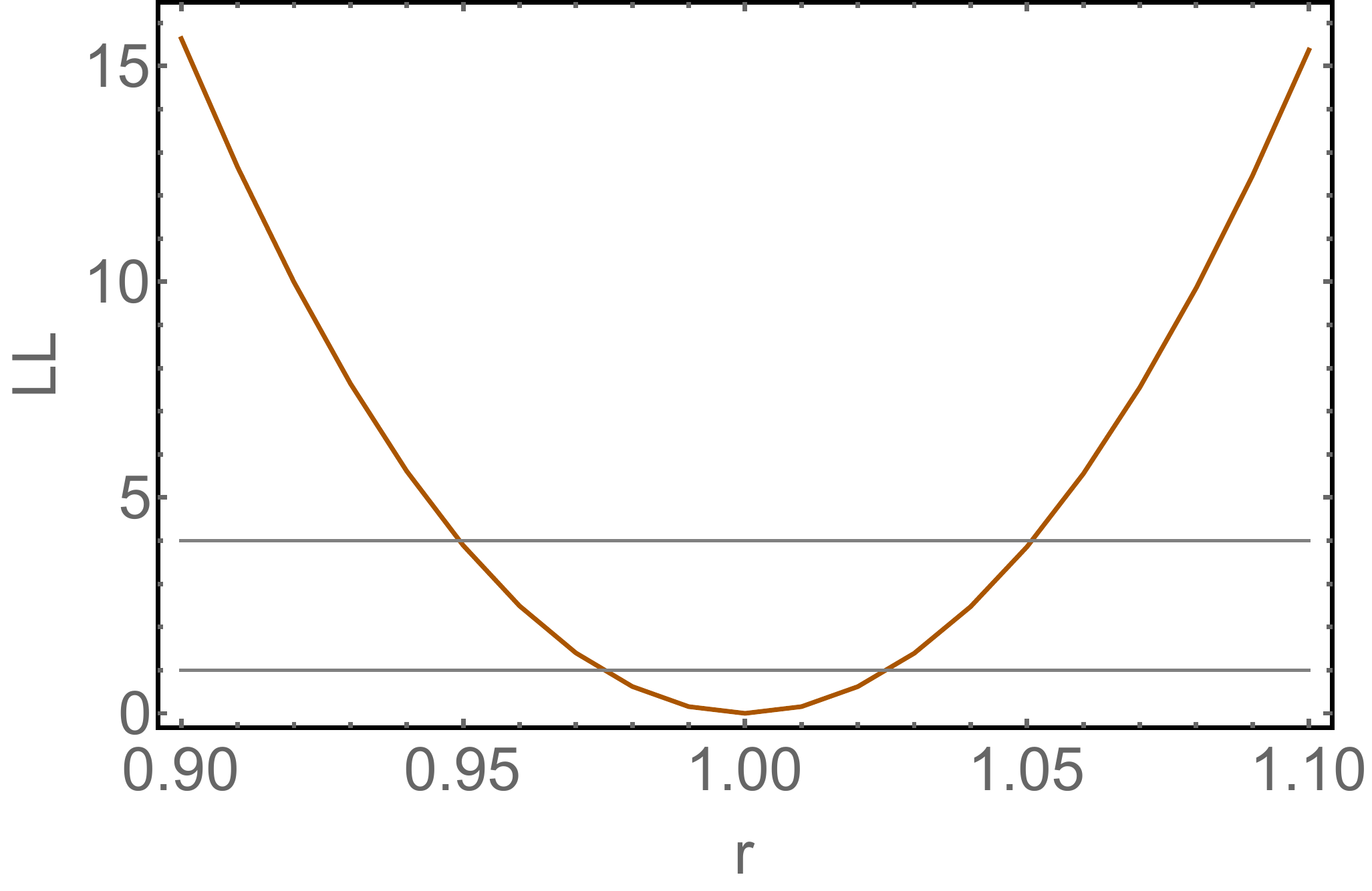}
	\caption[]{Five-dimensional Log likelihood as a function of the relative cross section $r$ defined below Eq.~(\ref{eq:LL}) 
	for the $500$ GeV case (left) and the $1$ TeV case (right). For both analyses, Cut 1 is applied.}
	\label{fig:ll500}
\end{figure}

To improve upon the cut-based results for reaching the optimal sensitivity, we perform a multivariate analysis (MVA) by evaluating a five-dimensional log-likelihood as a function of the deviation from the SM. Assuming Poisson statistics in each bin, the log-likelihood is defined as 
\begin{equation}
LL (\mathbf{n};\boldsymbol{\nu}) = 2 \sum\limits_{i=1}^{N_\text{bins}}  [\ n_i\ ln (\frac{n_i}{\nu_i}) + \nu_i - n_i ]
\label{eq:LL}
\end{equation}
where $\nu_i$ is the expected number of events in bin $i$ for the SM signal plus background, and $n_i$ is the number of events in bin $i$ for the SM signal scaled by factor $r$ (signal $\times$ $r$) plus background. We evaluate the region around $r=1$ and our $1\sigma$ deviation from the Standard Model value corresponds to $ \Delta LL = 1$.

Rather than applying Cut 2 on the angular distributions,
we apply Cut 1 and evaluate the log-likelihood in the five dimensional phase space defined by the variables $m_\text{rec}$, $m_{ee}$, $\theta_1$, $\theta_2$, and $\phi$. In the analysis, we perform a 3125-bin analysis by dividing the phase space along each variable evenly into five bins. Figure~\ref{fig:ll500} shows the log likelihood as a function of $r$. In the $500$ GeV analysis, we find the sensitivity on signal cross section improved to $6.0\%$. For the $1$ TeV case, the multivariate analysis increases the sensitivity to $2.5\%$. The likelihood profile for the $500$ GeV ($1$ TeV) case is shown in the left (right) panel of Fig.~\ref{fig:ll500}.

%%%%%%%%%%%%%%%%%%%%%%%%%%%%%%%%%%%%%%
\section{Impact on Higgs Physics}
\label{sec:disc}
\subsection{Higgs width and coupling Fits}
\label{sec:fit}

\begin{table}[]
  \centering
    \begin{tabular}{|c|c|c|c|c|}

    \hline
    \multirow{1}[0]{*}{Relative error \%} & \multicolumn{2}{|c|}{\multirow{1}[0]{*}{ILC 250+500}} & \multicolumn{2}{|c|}{\multirow{1}[0]{*}{ILC 250+500+1000}}  \\ \hline
     $\delta \sigma_{Zh}$ & \multicolumn{2}{|c|}{6.0\%} & \multicolumn{2}{|c|}{2.5\%} \\ \hline\hline
      Improvement      &   & With HL-LHC & & With HL-LHC\\ \hline

    $\Gamma$ & 4.8 $\rightarrow$ 4.7 & 4.8 $\rightarrow$ 4.6 & 4.5 $\rightarrow$ 3.7 & 4.5 $\rightarrow$ 3.7 \\ \hline
    $g_Z$  & 0.99 $\rightarrow$ 0.94 & 0.99 $\rightarrow$ 0.94 & 0.98 $\rightarrow$ 0.75 &  0.98 $\rightarrow$ 0.75\\ \hline
    $g_W$  & 1.1 $\rightarrow$ 1.1 & 1.1 $\rightarrow$ 1.1 & 1.1 $\rightarrow$ 0.89 & 1.1 $\rightarrow$ 0.88 \\ \hline
    $g_b$  & 1.5 $\rightarrow$ 1.5 & 1.5 $\rightarrow$ 1.5 & 1.3  $\rightarrow$ 1.2 & 1.3 $\rightarrow$ 1.1 \\ \hline
    \end{tabular}%
\caption[]{The improvement on selected coupling precisions by incorporating our $ZZ$ fusion analysis from a typical 10-parameter model-independent fit. We show both the ILC exclusive results and ILC combined with the optimistic CMS HL-LHC input~\cite{Dawson:2013bba}. For details of fitting scheme and combination scheme, see Ref.~\cite{Han:2013kya}. The results for ILC 250/500/1000~(\gev)\ assume 250/500/1000~$\fbi$ integrated luminosities.}
\label{tab:mi}
\end{table}

Based on our results, the sensitivities on $\sigma_z^\text{inc}$ which can be reached by studying the $ZZ$ fusion channel at $500\gev$ and $1\tev$ ILC are 6.0\% (8.2\%) and 2.5\% (3.1\%) based upon MVA (cut-based) analyses, respectively. 
In combination this yields a 2.3\% (2.9\%) combined uncertainty on $\sigma_z^\text{inc}$ from this production mode.

This is comparable to the current estimated precision of the ILC from studies of $Zh$ associate production~\cite{Miyamoto:2013zva} (that is, $\sigma_z^\text{inc}$ of $2.0\%$ achieved by combining 2.6\% and 3.0\% uncertainties from 250 GeV and 500 GeV~\cite{Asner:2013psa}). Thus, by combining the $ZZ$ fusion and $Zh$ measurements we estimate a final sensitivity $\sigma_z^\text{inc}$ to 1.5\%, a 25\% improvement over the $Zh$ channel alone. 
This improvement refines many other derived quantities in the model-independent fit. We demonstrate the improvement for a few representative quantities in Table~\ref{tab:mi}. We have performed a global 10-parameter model-independent fit following Ref.~\cite{Han:2013kya}. We compute sensitivities for the ILC alone and in combination with projected High Luminosity (HL)-LHC results. We take the optimistic projections for HL-LHC precision on cross sections from the CMS detector based on Ref.~\cite{Dawson:2013bba}.
As discussed in detail in Ref.~\cite{Han:2013kya}, twice the error of $\sigma_z^\text{inc}$ propagates into the $\Gamma_{\rm tot}$ determination, and this error dominates for stages beyond the 250~GeV phase of the ILC. Our study at the ILC 250+500+1000 stage relatively improves the total width precision by 16\%, Higgs to $ZZ$ coupling by 25\%, Higgs to $WW$ coupling by 16\%, and Higgs to $b\bar b$ coupling by 8\%. For other couplings with less precision the $\sigma_z^\text{inc}$ is not the largest source of uncertainty and less improvement is expected.

\subsection{Operator analysis}
\label{sec:opt}

New physics beyond the Standard Model (BSM) could give rise to modifications of the Higgs couplings. 
The proper framework to describe such possibilities in a model-independent manner is the effective field theory approach. With respect to the SM gauge symmetry, such effects are expressed by dimension-six Higgs operators after integrating out heavy particles or loop functions \cite{Weinberg:1978kz,Leung:1984ni,Buchmuller:1985jz,Hagiwara:1993qt}.\footnote{For recent reviews of these operators, see e.g., Refs.~\cite{Giudice:2007fh,Grzadkowski:2010es,Contino:2013kra,Agashe:2014kda}. Many of these operators not only contribute to Higgs physics, but also modify electroweak precision tests simultaneously~\cite{Ellis:2014dva,Ellis:2014jta,Biekoetter:2014jwa,Falkowski:2014tna}. }
The operators modifying Higgs to $ZZ$ couplings are naturally of particular interest in our case. This is partly because it will be one of the most precisely determined quantities through a recoil-mass measurement and partly because it is one of the key couplings that could help reveal the underlying dynamics of electroweak symmetry breaking. 
Certain operators may have different momentum dependence and thus measurements of differential cross sections may be more sensitive to the new effects.\footnote{For discussions of the effects on Higgs decays due to these operators, see Ref.~\cite{Beneke:2014sba}.}
The ILC is expected to have several operational stages with different center-of-mass energies, and the high-precision measurement achievable from $ZZ$ fusion will contribute to our knowledge of these different operators.\footnote{Assuming existence of a single operator at a time, limits can be derived, see, e.g.,~\cite{Craig:2014una}.}

To demonstrate this important feature, we consider the following two representative operators 
\begin{align}
\mathcal{O}_{H} = \partial^\mu( \phi^{\dagger}\phi) \partial_\mu( \phi^{\dagger}\phi ), \qquad 
\mathcal{O}_{HB} = g'  D^\mu\phi^{\dagger}D^\nu \phi B_{\mu\nu}, 
\label{eq:opt}
\end{align}
with
\beq
\mathcal{L}^{dim-6}\supset 
\frac{ c_H}{2\Lambda^2}\mathcal{O}_{H} + \frac{c_{HB} } {\Lambda^2} \mathcal{O}_{HB},
\label{eq:lag}
\eeq
where $\phi$ is the SM SU$(2)_L$ doublet and $\Lambda$ is the new physics scale. 
The coefficients $c_H$ and $c_{HB}$ are generically of order unity. 
Following the convention for comparison with existing studies \cite{Hagiwara:1993qt,Contino:2013kra,Ellis:2014dva,Ellis:2014jta,Biekoetter:2014jwa,Falkowski:2014tna}, we adopt the scaled coefficients 
$\bar c_H = \frac {v^2} {\Lambda^2} c_H$ and $\bar c_{HB} = \frac {m_W^2}{\Lambda^2} c_{HB}$.
This translates to generic values of $\bar c_H\approx0.06$ and $\bar c_{HB}\approx0.006$ for $\Lambda=1~\tev$.

The operator $\mathcal{O}_{H}$ modifies the Higgs-$ZZ$ coupling in a momentum-independent way at lowest order. This operator renormalizes the Higgs kinetic term and thus modifies the Higgs coupling to any particles universally \cite{BargerHanZerwasetal,Craig:2014aea}.
Equivalently, one may think of rescaling the standard model coupling constant. 
In contrast, the operator  $\mathcal{O}_{HB}$ generates a momentum-dependent Higgs-$ZZ$ coupling. This leads to a larger variation of the production rate versus c.m.~energy for the $Zh$ process than 
the $ZZ$ fusion because of the energy difference in the intermediate $Z$ bosons. Consequently, the corresponding deviations of the cross sections are approximately
\bea
{\rm ILC}~250~\gev:& \frac {\Delta \sigma} {\sigma}(Zh)&\approx-\bar c_H~-4.5~\bar c_{HB},\nonumber \\
{\rm ILC}~500~\gev:& \frac {\Delta \sigma} {\sigma}(Zh)&\approx-\bar c_H~- 25~\bar c_{HB},\nonumber \\
& \frac {\Delta \sigma} {\sigma}(e^{-} e^{+}h)&\approx-\bar c_H~+ 1.1~\bar c_{HB},\\
{\rm ILC}~~~~1~\tev:& \frac {\Delta \sigma} {\sigma}(e^{-} e^{+}h)&\approx-\bar c_H~+2.4~\bar c_{HB}.\nonumber
\eea

Such operators receive direct constraints from the LHC from similar production processes~\cite{Ellis:2014dva,Ellis:2014jta}, off-shell Higgs-to-$ZZ$ measurement~\cite{Azatov:2014jga}, etc., all of which lack desirable sensitivities due to the challenging hadron collider environment.
Based on an analysis of current data the coefficient $\bar c_{HB}$ is excluded for values outside the window $( -0.045, 0.075)$\footnote{The window is $(-0.053, 0.044)$ for single-operator analysis. This smallness of the difference between the marginalized analysis and single-operator analysis illustrates that this operator mainly affects Higgs physics and thus other electroweak precision observables do not provide much information.} and $\bar c_H$ is far less constrained~\cite{Ellis:2014dva,Ellis:2014jta}.

\begin{figure}[t]
	\centering
	\includegraphics[scale=.8]{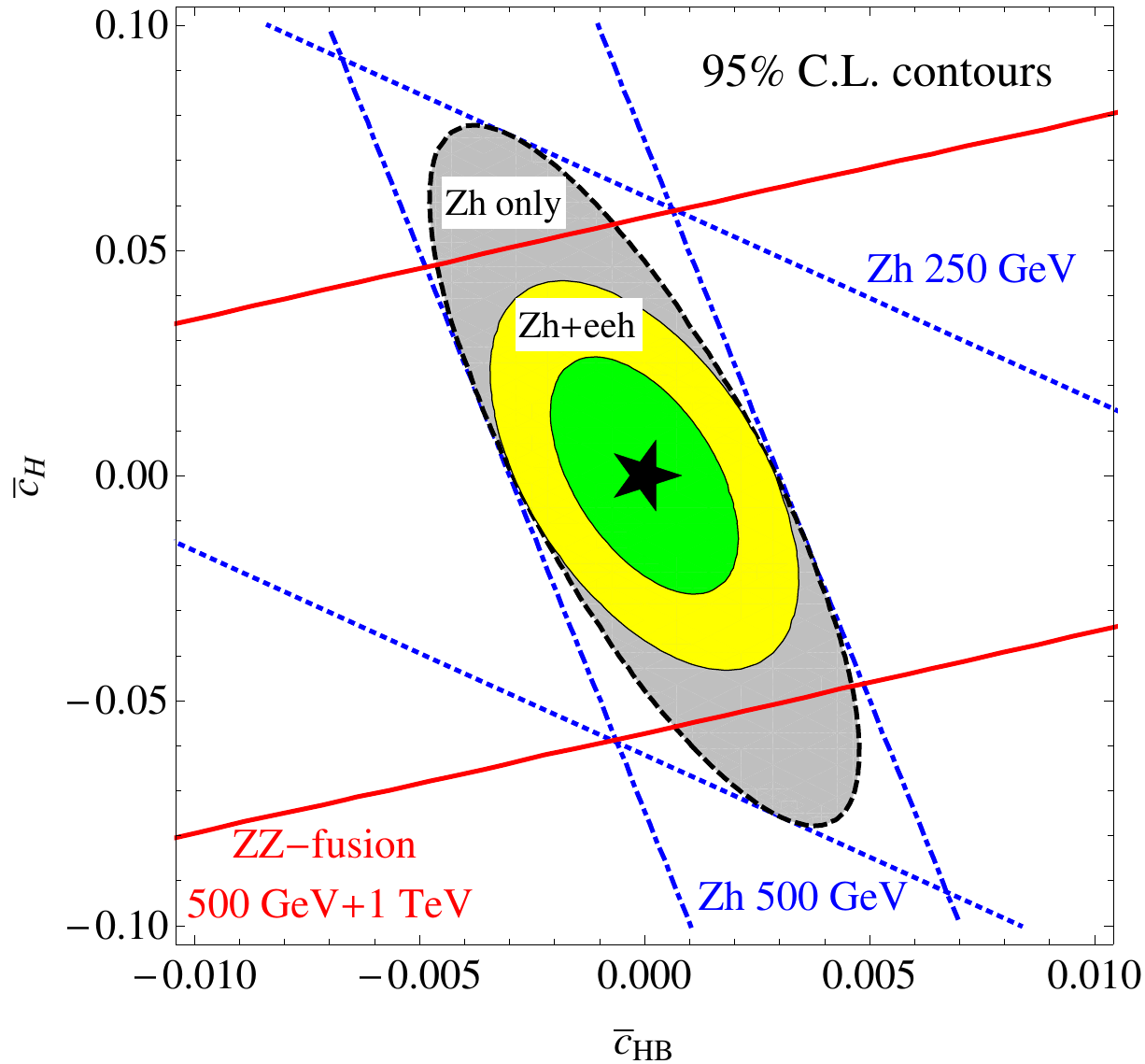}
	\caption[]{
	Constraints on coefficients of dimension-six operators $\bar c_H$ and $\bar c_{HB}$ with and without the inclusion of the $ZZ$ fusion channel. The dashed and dot-dashed lines represent $2\sigma$ deviations from zero in the $Zh$ channel at $250$ and $500$ GeV  (blue lines), respectively. The solid (red) lines indicates the constraint from $ZZ$ fusion for $500~\gev$ plus $1~\tev$. The outer (black-dashed) contour shows the constraint from combined $Zh$ measurements and the middle (yellow) and inner (green) contours show the combined $2\sigma$ and $1\sigma$ results with $ZZ$ fusion included.
}
\label{fig:contour}
\end{figure}

We only list above the cross sections which can be precisely measured at different ILC stages, with corresponding polarizations taken into account. The distinction between $ZZ$ fusion($e^{-} e^{+}h$) and $Zh$-associated production with $Z$ decaying to electron-positron pairs is easily made by applying a minimal $m_{ee}$ cut above $m_Z$.

In Fig.~\ref{fig:contour} we plot the expected constraints on the constants $\overline{c}_H$ and $\overline{c}_{HB}$ from the $Zh$ and $ZZ$ processes measured at the ILC, assuming only these two constants among the six-dimensional terms are nonzero. We show the 95\% C.L. contours for different measurements. The dashed(dot-dashed) blue line represents the contour from $Zh$-associated measurement at ILC 250~\gev (500~\gev). The red line represents the contour from combined $ZZ$ fusion measurements at ILC 500~\gev\ and 1~\tev. One can see that at a given energy for a simple production mode only a linear combination of the two operators is constrained, resulting in a flat direction in the contours. However, measurements of $Zh$ at two different energies would allow us to measure both simultaneously, as shown in the gray contour. Moreover, the addition of the $ZZ$ information at $1$ TeV would offer {\it significant} improvements as shown in the yellow contour. This allows us to measure $\overline{c}_H$ and $\overline{c}_{HB}$ at the level of $0.04$ and $0.004$ respectively. Much of the improvement comes from the fact that in $ZZ$ fusion, in contrast to $Zh$-associate production, the $\mathcal{O}_{HB}$ operator contributes with the opposite sign of the $\mathcal{O}_{H}$ operator.
We note here such indirect measurements would strongly constrain  BSM physics which  are otherwise difficult to test, such as singlet-Higgs assisted baryogenesis~\cite{Curtin:2014jma}, ``neutral naturalness''~\cite{Chacko:2005pe,Burdman:2006tz,Craig:2014aea}, etc.

\section{Conclusions}
\label{sec:conclusion}
To summarize, the $ZZ$ fusion channel for Higgs measurement could provide valuable information for  precision studies of the Higgs width and couplings because of the logarithmic increase of the total cross section versus the center-of-mass energy as seen in Fig.~\ref{fig:xsec}. 
Although the signal suffers from large radiation-induced smearing at high energies it can be observed with good precision at a $1$ TeV run and benefits from a multivariate analysis. We have also demonstrated the sensitivity to probe higher-dimensional operators at the ILC, which are usually not  covered by conventional global fits.  
We find:

(i)
The inclusive cross section of the $ZZ$ fusion channel can be measured to $2.5\%$ at $1$ TeV.  This is competitive with the best estimate of Higgsstrahlung measurement at $250$ GeV, as shown in Secs.~\ref{sec:1TeV} and \ref{sec:LL}. 

(ii)
Combing the $ZZ$ fusion and Higgsstrahlung channels,  
the model-independent measurement of the inclusive cross section can be improved to $1.5\%$ with a commensurate improvement of the Higgs 
width determination, as shown in Sec.~\ref{sec:fit}.

(iii)
Sensitivities on the inclusive cross section $\sigma_{Z}^\text{inc}$ at multiple energies also offer the possibility to distinguish contributions from different higher-dimensional operators induced by BSM physics. We demonstrate the ability to simultaneously constrain two operators whose effects are difficult to observe at the LHC, as shown in Sec.~\ref{sec:opt}.
Including the $ZZ$ fusion channel provides as large as $50\%$ relative improvement for the constraint on the chosen operators compared to the $Zh$-associated production channel alone.

In the preceding analysis and discussion, we have shown the appreciable impact of including the $ZZ$ fusion channel at the ILC for Higgs physics. Full detector simulations may be desirable to further the study of this signal mode.

\acknowledgments{}

This work was supported in part by the U.S.~Department of Energy under Grant
No.~DE-FG02-95ER40896 and in part by the PITT PACC. Z.\,L. was also supported in part by the Andrew Mellon Predoctoral Fellowship and a PITT PACC Predoctoral Fellowship from Dietrich School of Art and Science, University of Pittsburgh, and in part by the Fermilab Graduate Student Research Program in Theoretical Physics.

\appendix

\section{Consideration of one-photon sensitivity}

As discussed in the main text, we find it useful to simply veto events with a single, isolated photon in addition to an electron-positron pair. This cut reduces the potentially large background arising from Bhabha scattering plus radiation which can pass the invariant mass and $p_T$ cuts. This cut also reduces signal events where the Higgs decays to a single photon plus invisible particles, or a single photon plus additional particles which are lost down the beam pipe. In general we do not expect this to be a relevant effect since our final sensitivity for the model-independent cross section is $2.5\%$ while the Standard Model processes which might contribute to such events are at the level of $10^{-3}$ branching fractions or less. Only order of magnitude enhancements to these channels from exotic physics would be relevant to our analysis and such enhancements are constrained by  exclusive searches at the LHC and in future at the ILC.

Nevertheless, there may be some exotic model which would produce an observable effect in the inclusive measurement which is not ruled out by other searches. We note that if one wishes to preserve sensitivity to exotic channels which could produce a single isolated photon, it is possible to institute cuts which will remove almost all of the background while preserving a substantial fraction of any such Higgs decays. We find that, in the reconstructed Higgs rest frame, the isolated photon in the background sample is not isotropically distributed. The background photon usually appears collinear to 
the Higgs boost direction, and/or confined to be near the radial plane containing the beam and the Higgs boost vector. This is because the photon is recoiling against the $e^-e^+$ pair with a possible boost along the beam axis due to additional unseen photons. We also find that measurement errors on the photon are typically larger in the polar angle than in the azimuthal direction. Thus one can largely remove this background by cutting on the polar (with respect to the Higgs boost) and azimuthal (measured with respect to the Higgs-beam plane) angles of a single extra photon in the Higgs rest frame.  We find the problematic background can be reduced to the level of a few fb while preserving $\sim 60 \%$ of any hypothetical Higgs decay signal, \footnote{This fraction is relative to other decay channels not affected by the cut, since other cuts will affect all decays equally.} since the photon from such a decay would be isotropically distributed in the Higgs rest frame. Hence any new physics signal large enough to affect the inclusive rate would still be observable, although underestimated. 

We note that a cut similar in spirit to this one is already present in the widely used analysis of Higgsstrahlung-inclusive measurement at the $250$ GeV ILC \cite{Li:2010wu}. In that case additional single photons were removed by a ``$p_T$ balance" cut when the $p_T$ of an isolated photon accounted for the bulk of the $e^-e^+$ pair $p_T$.  However, since this more complicated approach does not materially change our results we  present the simpler case of simply vetoing the single isolated photon as described in the main text.

\bibliographystyle{jhep}
\bibliography{references}

\end{document}